\documentclass[pre,aps,floatfix,notitlepage]{revtex4-1}

\usepackage[dvips]{graphicx}
\usepackage{amssymb,amsfonts,amsmath,calc,ifsym,bbding,wasysym}
\usepackage[usenames]{color}
\usepackage{bm}
\usepackage[normalem]{ulem}
\usepackage{xr}
\usepackage{xspace}
\usepackage{float}
\usepackage{grfext}
\usepackage{braket}
\usepackage[usenames,dvipsnames]{xcolor}
\usepackage{siunitx}

\PrependGraphicsExtensions*{.pdf,.PDF}  

\definecolor{gold(metallic)}{rgb}{0.83, 0.69, 0.22}

\newcommand{\eg}{\textit{e.g.}\xspace}
\newcommand{\invitro}{\textit{in vitro}\xspace}
\newcommand{\invivo}{\textit{in vivo}\xspace}

\def\ld{\lambda_\text{D}}
\def\Csalt{C_\text{salt}}
\def\I{I}
\def\ess{\varepsilon_\text{ss}}
\def\eps{\varepsilon_\text{PS}}
\def\nps{N_\text{PS}}
\def\gss{g_\text{ss}}
\def\gps{g_\text{PS}}
\newcommand{\kd}{K_\mathrm{d}}
\newcommand{\sbb}{s_\mathrm{b}}

\def\Lopt{L^{*}_\text{eq}}

\newcommand{\kt}{k_\mathrm{B}T}
\def\eunit{E_\text{u}}

\def\dunit{D_\text{u}}

\def\qp{q_\text{p}}

\def\rcut{r_\text{cut}}
\newcommand{\sub}[2]{ _{\mathrm{#1}#2}}
\newcommand{\rsub}[1]{_\mathrm{#1}}

\newcommand{\LJ}[1]{ \mathcal{L}_{#1} }
\newcommand{\Morse}[1]{ \mathcal{M}_{#1} }

\newcommand{\nfree}{n_\text{free}}

\newcommand{\rstoich}{r_\text{ex}}

\newcommand{\rh}{R_\text{H}}

\newcommand{\bad}{\langle \Delta B(n)\rangle}
\def\tend{t_\text{end}}

\begin{document}

\title{The role of packaging sites in efficient and specific virus assembly}
\author{Jason D Perlmutter}
\author{Michael F Hagan}
\email{hagan@brandeis.edu}
\affiliation{Martin Fisher School of Physics, Brandeis University, Waltham, MA, USA.}
\maketitle

\section{Abstract}

During the lifecycle of many single-stranded RNA viruses, including many human pathogens, a protein shell called the capsid spontaneously assembles around the viral genome.  Understanding the mechanisms by which capsid proteins selectively assemble around the viral RNA amidst diverse host RNAs is a key question in virology. In one proposed mechanism, sequence elements (packaging sites) within the genomic RNA promote rapid and efficient assembly through specific interactions with the capsid proteins.  In this work we develop a coarse-grained particle-based computational model for capsid proteins and RNA which represents protein-RNA interactions arising both from nonspecific electrostatics and specific packaging sites interactions. Using Brownian dynamics simulations, we explore how the efficiency and specificity of assembly depend on solution conditions (which control protein-protein and nonspecific protein-RNA interactions) as well as the strength and number of packaging sites. We identify distinct regions in parameter space in which packaging sites lead to highly specific assembly via different mechanisms, and others in which packaging sites lead to kinetic traps. We relate these computational predictions to \invitro assays for specificity in which cognate viral RNAs are compete against non-cognate RNAs for assembly by capsid proteins.

\pagebreak
\section{Introduction}
In many single-stranded RNA virus families, the spontaneous assembly of a protein container (capsid) around the viral RNA is an essential step in the viral life cycle \cite{Hagan2014}. Formation of an infectious virion requires that the assembling proteins select the viral RNA out of the milieu of cellular RNA,  and most viruses do so with high specificity (e.g. 99\% \cite{Routh2012}) \invivo. Understanding the mechanisms which enable such specific co-assembly could guide the design of delivery vectors that assemble around specific drugs or genes, and could identify targets for antiviral agents that interfere with genome packaging. In this work, we use dynamical computer simulations to investigate the ability of sequence-specific RNA-protein interactions (packaging sites) to drive specific packaging of the viral genome, and how specificity depends on the underlying sequence-independent interactions.

A key driving force for RNA-capsid protein co-assembly is provided by electrostatic interactions between RNA phosphate groups and basic amino acids, often located in flexible tails known as arginine rich motifs (ARMs) (e.g., \cite{ Schneemann2006}).  These nonspecific interactions are sufficient for assembly, as shown by the ability of capsid proteins to assemble \emph{in vitro} around heterologous RNA, synthetic polyelectrolytes, and other negatively charged substrates \cite{Bancroft1969,Hohn1969,Chen2005,Sun2007,Hu2008d,Sikkema2007,Brasch2012,Kostiainen2011,Goicochea2007,Loo2007,Kwak2010,Chang2008,Malyutin2013,Cheng2013}.  \textit{In vitro} assembly assays \cite{Comas-Garcia2012} and computational modeling \cite{Perlmutter2013,Erdemci-Tandogan2014} indicate that the charge and structure arising from base pairing of viral RNAs is optimal for assembly by their capsid proteins. However, these physical characteristics alone cannot explain the remarkably specific packaging of the viral genome achieved by many RNA viruses \textit{in vivo}. Several factors have been proposed to explain specific packaging \textit{in vivo}, including subcellular localization of viral components \cite{Bamunusinghe2011}, coordinated translation and assembly \cite{Dykeman2014, Annamalai2008,Kao2011}, and NA-sequence-specific interactions between capsid proteins and sites within the genome called packaging sites (PSs).  PSs have been identified for a number of unrelated viruses infecting plant, animal, or bacterial hosts, suggesting this mechanism has widespread relevance \cite{Rao2006,Stockley2013a,Pappalardo1998,Lu2011,D'Souza2005,Sorger1986,Beckett1988,Bunka2011,Dykeman2011}.

The specificity conferred by PSs has been explored through \textit{in vitro} experiments, either by comparing assembly yields of capsid proteins around cognate and non-cognate RNAs in separate experiments or by competition assays, in which two RNA species compete for packaging under limiting protein concentrations.  Measured selectivities have varied widely, ranging from high selectivity for the cognate \cite{Ling1970,Sorger1986,Beckett1988}, no selectivity \cite{Porterfield2010}, or selectivity for a non-cognate RNA\cite{Comas-Garcia2012}. Two recent experiments observed that assembly around cognate RNAs proceeded via different, faster assembly pathways than around non-cognate RNAs \cite{Borodavka2012,Ford2013}. The authors suggest that their experiments are more selective for cognate RNAs because they use a lower protein concentration than previous experiments ($1 \mu M$ vs $10 \mu M$). 

Using chemical kinetics simulations (Gillespie algorithm \cite{Bortz1975,Gillespie1977,Schwartz1998}), Dykeman et al. \cite{Dykeman2013a, Dykeman2014} predicted assembly under dynamic subunit concentrations, i.e. the concentration increase (`ramp') that occurs during an infection cycle in E. coli, could lead to 100\% specificity for RNAs with PSs (represented by nonuniform protein binding affinities) even under a large excess of non-cognate RNAs (represented by uniform binding affinities). In contrast, constant subunit concentrations led to weak differences in yield ($\sim 5\%$) and a significant portion of malformed capsids. However, these simulation results do not entirely address the recent \textit{in vitro} experiments \cite{Borodavka2012,Ford2013} in which PSs led to high yield assembly while non-cognate assembly was unsuccessful using constant subunit concentrations. A limitation of Gillespie algorithm simulations is that the state space (the set of allowed partial capsid geometries and RNA configurations) and the transition rates (e.g. association rates among RNA-bound subunits) must be assumed \textit{a priori} \cite{Hagan2014}. It is therefore difficult to account for complex processes such as cooperative RNA-protein motions seen in previous Brownian dynamics simulations \cite{Elrad2010,Kivenson2010}. While these assumptions can be guided by experimental data in certain cases, we seek here to determine the ensemble of possible assembly pathways and products.

We recently developed a particle-based computational model for RNA and capsid proteins \cite{Perlmutter2013,Perlmutter2014} with which capsid assembly is simulated using Brownian dynamics. Although the model is coarse-grained, model predictions for RNA lengths that optimize capsid thermostability quantitatively agreed with viral genome length for seven viruses \cite{Perlmutter2013}. We previously examined how varying the nonspecific electrostatic RNA-protein subunit interactions, solution conditions, and subunit-subunit interactions leads to a range of assembly outcomes and different classes of assembly pathways \cite{Perlmutter2014}.

Here, we explore how introducing specific PS interactions, in a simple form inspired by a recent structural investigation of STNV \cite{Ford2013}, alters these assembly pathways and products.  By extensively comparing assembly around uniform polyelectrolytes (representing non-cognate RNA) and PS-containing polyelectrolytes (cognate RNA), we identify solution conditions that lead to highly specific packaging of the cognate RNA. Depending on the relative strength of protein-protein and protein-RNA interactions, we find that PSs can drive specific assembly via several mechanisms.  Consistent with recent single molecule experiments \cite{Borodavka2012}, the simulations indicate that PSs can trigger assembly via pathways with more compact intermediates as compared to non-cognate RNAs. However, we also find solution conditions under which PSs are unable to drive specific packaging or even lead to kinetic traps. We then investigate how assembly yields and specificity depend on the number and strength of PSs. In general, we find that a combination of one high affinity PS and multiple weak PSs leads to the highest assembly yields, consistent with the identification of multiple weak PSs in viral genomes \cite{Stockley2013a} and with previous observations that productive self-assembly reactions require reversible interactions \cite{Whitelam2014, Hagan2014}.  We conclude by discussing potential experimental predictions suggested by these simulations.

\section{Model}
\label{sec:model}
To study the effect of PSs on assembly, we have extended a recently developed model \cite{Perlmutter2013,Perlmutter2014} for assembly around linear polyelectrolytes and non-cognate RNAs to include a representation of PSs.  The model is motivated by recent experiments in which purified simian virus 40 (SV40) capsid proteins assemble \textit{in vitro} around ssRNA molecules to form virus-like particles composed of 12 homopentamer subunits \cite{Kler2012,Kler2013}.  The model capsid is therefore a dodecahedron comprising 12 pentagonal subunits, each of which represents a homopentamer of the capsid protein.  It is assumed that homopentamers are stable and form rapidly in solution, as is the case for for SV40. Although the structure of the model capsid is motivated by these experiments \cite{Kler2012, Kler2013}, in this article we use the model to study general relationships between PSs and assembly which could apply to many viral species.

\begin{figure}[hbt]
\centering{\includegraphics[width=0.75\columnwidth]{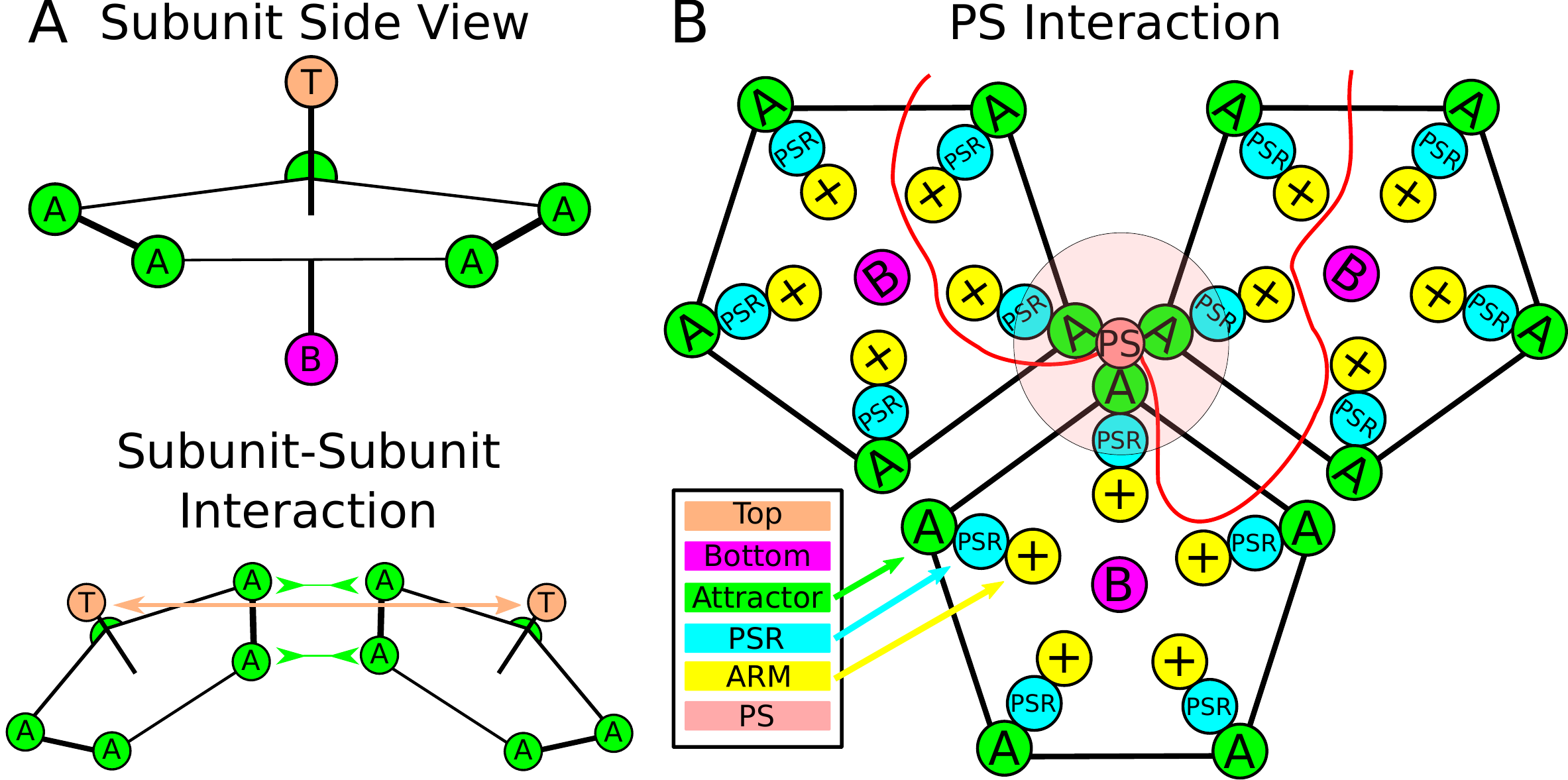}}
\caption{{\bf (A)} Model schematic showing components responsible for subunit-subunit interactions: subunits are bound together by attractor pseudoatoms (`A'), and the Top (`T') and Bottom (`B') pseudoatoms guide the subunits towards the correct geometry (see SI). {\bf (B)} Schematic with components responsible for attractive interaction with the RNA (drawn in red) and packaging site (`PS'): positively charged ARM (`+') and PS Receptor (`PSR'). The `Excluder' pseudoatoms, which represent the excluded volume of the capsid shell, are located within the black pentagons; to aid visibility, they are not explicitly drawn here.  Snapshots here and throughout the article are colored as follows: blue=excluders, green=attractors, yellow=ARM, red=RNA, orange=PS.}
\label{schematic}
\end{figure}

{\bf Capsid protein subunit-subunit interactions.} Following Refs \cite{Wales2005,Perlmutter2013,Perlmutter2014}, model subunits are attracted to each other via attractive pseudoatoms, `attractors' (type `A') at the vertices, which interact via a Morse potential  (see Fig.~\ref{schematic} and the Methods section).  The subunit-subunit interaction strength is controlled by the model parameter $\ess$; the free energy of subunit dimerization is $\gss/\kt{=}5.0-1.5\ess$. This does not include the effects due to repulsions between ARMs (defined below), which we estimate to reduce $\gss$ by $\sim0.5\kt$ at 100mM; see SI section~\ref{sec:freeenergy}. These attractions represent the interactions between capsid protein subunits that arise from hydrophobic, van der Waals, and electrostatic interactions \cite{Hagan2014}), whose strength can be experimentally tuned by pH and salt concentration \cite{Ceres2002,Kegel2004,Hagan2014}.

Pairs of subunits are driven toward a preferred subunit-subunit angle consistent with a dodecahedron (116 degrees) by repulsive `Top' pseudoatoms (type `T'), which interact via the repulsive term of the Lennard-Jones (LJ) potential. The `Bottom' pseudoatoms (type `B') have a repulsive LJ interaction with `T' pseudoatoms, intended to prevent `upside-down' assembly. The `T', `B', and `A' pseudoatoms form a rigid body \cite{Wales2005, Fejer2009, Johnston2010}. See Refs. \cite{Schwartz1998, Hagan2006, Hicks2006, Nguyen2007, Wilber2007, Nguyen2008, Nguyen2009, Johnston2010, Wilber2009a, Wilber2009, Rapaport1999, Rapaport2004, Rapaport2008, Hagan2008, Elrad2010, Hagan2011, Mahalik2012, Levandovsky2009} for related models.

{\bf Sequence-independent electrostatic interactions.}  Capsid assembly around nucleic acids and other polyelectrolytes is driven by electrostatic interactions between negative charges on the encapsulated polyelectrolyte and positive charges on capsid protein-RNA binding domains \cite{Schneemann2006,Hagan2014}.To account for these interactions, we extend the model as follows. First, we add positively charged bead-spring polymers affixed to the inner surface of the subunit, to represent the highly charged, flexible terminal tails known as arginine rich motifs (ARMs) that are typical of positive-sense ssRNA protein-RNA binding domains (e.g., \cite{Schneemann2006}). There are five ARMs per pentameric subunit. For each ARM, the first segment is anchored at a fixed position on the subunit, midway between the subunit center and a vertex. Except where stated otherwise, each ARM contains five segments of charge $+e$. To better represent the capsid shell, we include a layer of `Excluder' pseudoatoms, which have a repulsive LJ interaction with the RNA and the ARMs. The `Excluders' and first ARM segment are part of the subunit rigid body. ARM beads interact through repulsive LJ interactions and, if charged, electrostatic interactions.

To represent an RNA molecule, we consider a linear bead-spring polyelectrolyte, with a charge of -$e$ per bead and a persistence length comparable to that of ssRNA in the absence of base pairing. To focus on the effect of PSs, we do not consider RNA base pairing in this work; the effect of base pairing on assembly was considered in Ref.~\cite{Perlmutter2013}. We also previously determined how assembly depends on polyelectrolyte length ~\cite{Perlmutter2013}. In the present work, for each simulated salt concentration and capsid structure, we use the polyelectrolyte length that optimizes assembly. The values of optimal lengths as a function of salt concentration are shown in Fig.~\ref{length}.

Electrostatics are modeled using Debye-H\"{u}ckel (DH) interactions, where the Debye screening length ($\ld$) is given by $\ld \approx 0.3/\Csalt^{1/2}$ with $\ld$ in nm $\Csalt$ the concentration of monovalent salt in molar units.   Perlmutter et al. \cite{Perlmutter2013} showed that DH interactions compare well to simulations with explicit counterions for the parameter values under consideration; comparisons between simulations with DH interactions and those with explicit counterions are presented in Refs.~\cite{Perlmutter2013,Perlmutter2014} and Fig.~\ref{length}.

{\bf Simulations and units.} Simulations were performed with the Brownian Dynamics algorithm of HOOMD, which uses the Langevin equation to evolve positions and rigid body orientations in time \cite{Anderson2008, nguyen2011, LeBard2012}. Simulations were run using a set of fundamental units. The fundamental energy unit is selected to be $\eunit\equiv1 \kt$. The unit of length $\dunit$ is set to the circumradius of a pentagonal subunit, which is taken to be $1\dunit\equiv5$ nm so that the dodecahedron inradius of $1.46\dunit=7.3$ nm gives an interior volume consistent with that of the smallest $T{=}1$ capsids. To calculate the thermodynamic optimal encapsidation length, we placed a very long polymer in or near a preassembled capsid, with one of the capsid subunits made permeable to the polymer and performed unbiased Brownian dynamics. Once the amount of packaged polymer reached equilibrium, the thermodynamic optimum length $\Lopt$ was measured. We previously \cite{Perlmutter2013} found that this strategy closely matched that produced using the Widom insertion method \cite{Widom1963} as applied to growing polymer chains \cite{Kumar1991,Elrad2010}. Assembly simulations were run at least 10 times for each set of parameters, each of which were concluded at either completion, persistent malformation, or $2\times10^8$ time steps. This observation time was chosen based on the time after which assembly yields and outcomes change only logarithmically with time for most parameter values.  For all dynamics simulations there were 60 subunits with $\mbox{box size} {=} 200 \times 200 \times 200$ nm, resulting in a concentration of  $12 \mu$M.

{\bf Packaging sites (PSs).} Structures of PSs obtained for a number of viruses through x-ray crystal structures and/or bioinformatics correspond to short stem loops \cite{Stockley2013a}. For example, multiple short stem-loops with a single-stranded
loop motif of –A.X.X.A-, where X corresponds to any nucleotide, were identified as PSs in the STNV genome \cite{Ford2013}.  An x-ray structure of STNV VLPs containing stem loop fragments revealed that they bind to well-defined sites on the protein ARMs with the effect of bringing multiple subunits into proximity and favoring alignments conducive to subunit-subunit interaction, thus enhancing subunit-subunit interactions as well as subunit-RNA interactions \cite{Ford2013}.

To account for the these effects, we have extended the model to include a generic representation of PS interactions by adding new pseudoatoms (denoted as packaging site receptors, PSRs) to the model protein subunits (Fig.~\ref{schematic}B). The PSRs experience short-range interactions with particular RNA segments that correspond to PSs. For simplicity, the PS-PSR interaction uses the same short-range Morse potential as the attractor-attractor interaction (Eq.~\ref{eq:Morse}, section~\ref{sec:potentials}). The strength of the PS-PSR attractive interaction is parameterized by the interaction well-depth $\eps$.   There are five PSRs per pentameric subunit. Except where noted otherwise, each PSR is located approximately midway between an ARM anchor segment and a subunit vertex.  This location allows for a PS to simultaneously bind to 3 PSRs when three subunits form an optimal configuration. Thus, the PSs not only promote subunit-RNA binding, but also generate RNA-mediated subunit-subunit interactions, as inferred from structural data \cite{Ford2013}.

Recent experiments have identified multiple PSs within several viral genomes (\eg \cite{Bunka2011,Ford2013,archer2013, athavale2013, Schroeder2011, Zeng2012}) and that these PSs bind capsid proteins with a range of affinities \cite{Bunka2011,Dykeman2013b}. Typically there are one or a few high affinity PSs (e.g. nM $K_\text{D}$), with the remainder having weaker affinities (up to $\mu$M $K_\text{D}$) \cite{Dykeman2013b}.

In our simulations, we explore how assembly depends on \textbf{(A)} the number of PS ($\nps$), \textbf{(B)} the PS binding affinity ($\eps$), and \textbf{(C)} distribution of PS binding affinities along the polyelectrolye. \textbf{(A)} There are 60 PSRs in a complete model capsid, located at 20 threefold axes.  Thus, an RNA with 20 PSs can interact with every PSR, and we will refer to $\nps=20$ as the `stoichiometric' number of PSs. \textbf{(B)} To limit the number of model parameters, we consider two classes of model PSs:  high affinity sites, with PS-PSR interaction well depth $\eps=20\kt$ (see Eq~\ref{Usp}) and low affinity sites, with $\eps=5\kt$ (see Section~\ref{sec:freeenergy} for discussion of binding free energies). \textbf{(C)} To describe how assembly depends on the distribution of affinities, we consider three forms of PS distributions along the model RNA:  (\textit{i}) $\nps$ high affinity PSs, (\textit{ii}) $\nps$ low affinity PSs, and (\textit{iii}) 1 high affinity PS along with $\nps$ low affinity PSs.  In each case the PSs are placed at a uniform interval along the RNA. Unless otherwise noted, the strong PS in distribution \textit{iii} is placed at the center of the RNA as found for MS2 \cite{Beckett1988}.

{\bf Comparison to existing models.}
Dykeman et al. \cite{Dykeman2013a, Dykeman2014} extended the kinetic rate equation approach of Becker and Doring \cite{Becker1935} and Zlotnick \cite{Endres2002} to include a representation of RNA and PSs.  In this model the Gillespie algorithm \cite{Bortz1975,Gillespie1977,Schwartz1998} is used to stochastically sample paths according to a predefined state space and matrix of inter-state transition rates. They assume that each of the PSRs is bound by RNA once and that subunits adding to partial capsids are bound to adjacent RNA segments, so that assembly must follow a Hamiltonian path.
 In the present model the spatial positions and dynamics of subunits are explicitly tracked and thus there are no assumptions made about the state space or assembly pathways. Consequently, there are no explicit restrictions on the sequence of RNA binding sites, although steric hindrances disfavor binding of multiple PSs at the same threefold axis and RNA conformational statistics favor returning to nearby binding sites.

\section{Results}
\label{sec:results}
In this section we describe how assembly depends on the specific (PS-PSR) and nonspecific (electrostatic) interactions. We refer to polyelectrolytes equipped with only nonspecific interactions as \textbf{non-cognate} RNAs, and polyelectrolytes which contain one or more PSs as \textbf{cognate} RNAs. Note that since we neglect base pairing in this work, the non-cognate RNA is a linear polyelectrolyte.  However, since we also neglect base pairing in the cognate RNA, our results are applicable to a comparison between cognate and non-cognate RNAs, except to the extent that the tertiary structure of cognate RNAs is more favorable for assembly than that of non-cognate RNAs ~\cite{Perlmutter2013, Yoffe2008}.

\begin{figure}[h]
\centering{\includegraphics[width=0.75\columnwidth]{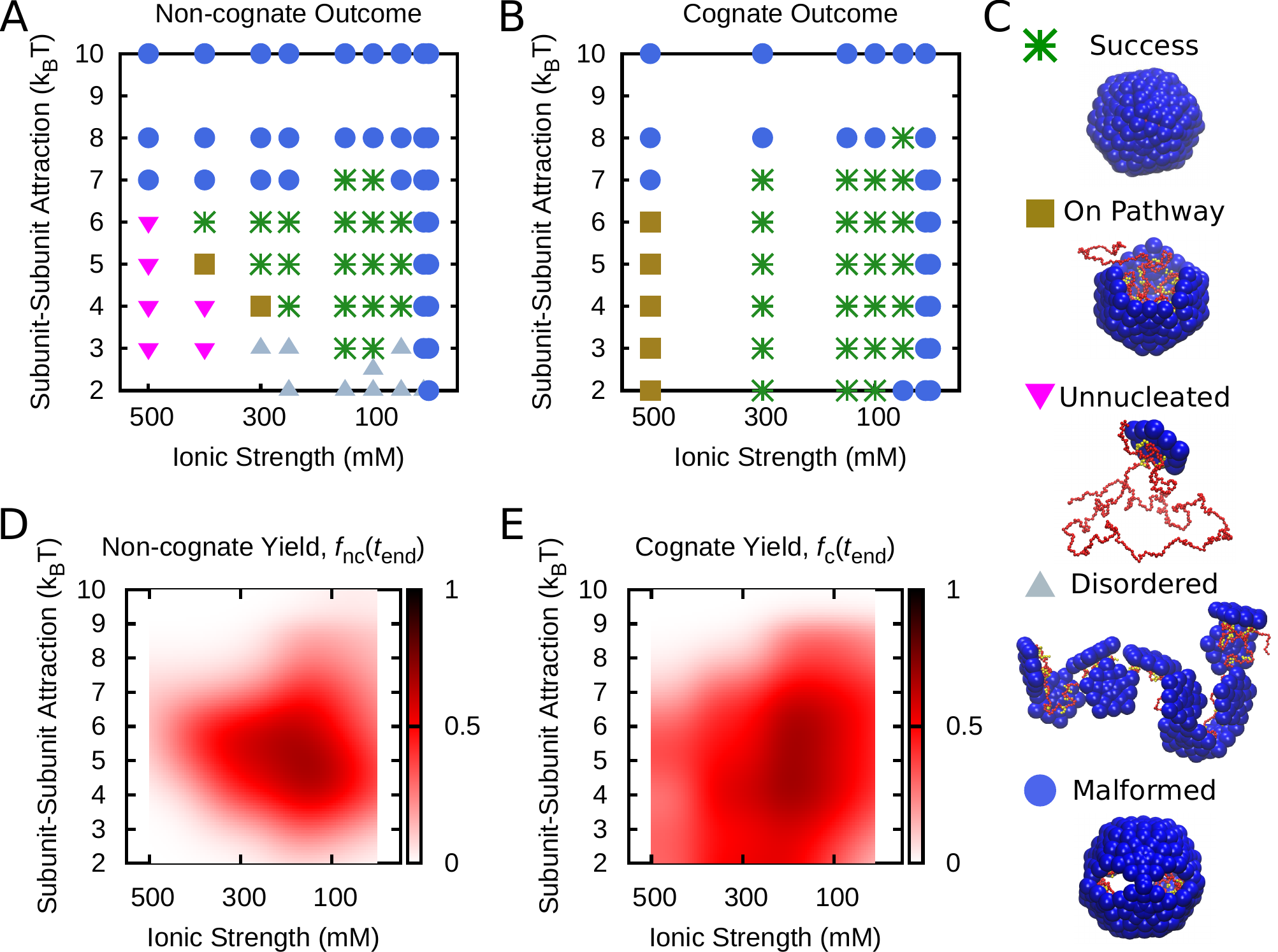}}
\caption{The effect of PSs and solution conditions on assembly yields and products. {\bf (A,B)}The most prevalent assembly product is shown as a function of ionic strength $\Csalt$ and subunit-subunit attraction well-depth $\ess$ for assembly around (A) a non-cognate RNA (polyelectrolyte without PS), and (B) a cognate RNA with 1 high affinity (HA, $\eps=20 \kt$) PS and 25 low affinity (LA, $\eps=5\kt$) PSs (B). A legend showing the outcome and a representative simulation snapshot corresponding to each symbol is presented in {\bf (C)}. {\bf (D,E)} The yield of well-formed capsids assembled around  (D) the non-cognate RNA or (E) the cognate RNA with the PS sequence as in (B). In each simulation the RNA length corresponds to the thermodynamic optimal length for the non-cognate at the simulated value of $\Csalt$, and ranges from 350 to 575 RNA segments} (see Fig.~\ref{equilibrium}).
\label{relativeYield}
\end{figure}

\subsection{The yield and specificity conferred by PSs depends on subunit-subunit and nonspecific electrostatic interactions}
\label{sec:specificity}
{\bf Yield without PSs.} The assembly of non-cognate RNA (uniform polyelectrolytes) depends on the strength of subunit-subunit interactions (controlled by $\ess$ in our model, eq.~\ref{Uss} in section~\ref{sec:potentials}, and salt concentration or $p$H \textit{in vitro}) and sequence-independent electrostatic interactions (controlled by the salt concentration $\Csalt$).  The dependence of assembly outcomes on these parameters for a non-cognate RNA is summarized in Fig.~\ref{relativeYield}.  High yields of well-formed VLPs are observed for  $\Csalt \in [50-400]$ mM and moderate subunit-subunit interaction strengths, $\ess \in [4-6]\kt$. Outside of optimal parameter values, yields are suppressed by several failure modes: strong electrostatics lead to disordered aggregates, strong subunit-subunit interactions lead to malformed capsids, and overly weak interactions lead to unnucleated complexes \cite{Perlmutter2014}. The yield $f_\text{nc}(t)$ is defined as the fraction of simulations which, at time $t$, resulted in formation of a complete capsid (defined as 12 subunits each strongly interacting with five neighbors) completely encapsulating the non-cognate RNA. The yield $f_\text{nc}(\tend)$ at the simulation endpoint $\tend$ is shown in Fig.~\ref{relativeYield}D. For each case, at least 10 simulations are run, so the estimated error in the yield ranges from $0.07-0.13$ \cite{Agresti1998}.

{\bf Yield with PSs.} Based on observations of multiple low affinity PSs \cite{Stockley2013a} and simulations at varying numbers and strengths of PSs (section~\ref{sec:number} below), we performed simulations at varying $\ess$ and $\Csalt$ for an RNA with 1 high affinity PS and $\nps{=}25$ low affinity PSs (see section~\ref{sec:potentials}). With the addition of PSs, the range of parameters leading to high assembly yields ($f_\text{c}$) broadens considerably (Fig.~\ref{relativeYield}B,E), allowing assembly at much lower values of $\ess$ across a wide range of $\Csalt$ and increasing the upper range of $\ess$ leading to assembly at low $\Csalt$.

Notably, assembly around the non-cognate RNA fails in distinct ways in these parameter regions (Fig.~\ref{relativeYield}C), indicating that PSs can avoid multiple forms of thermodynamic or kinetic traps. At low $\ess$, the predominant effect of PSs is to enhance nucleation and growth rates  by increasing effective interaction strengths.  Increased subunit-RNA interactions are most relevant at high salt while increased subunit-subunit interactions are most relevant at low salt (discussed in section~\ref{sec:number}).  At high $\ess$  and low or moderate salt (\eg $\Csalt{=}50$mM and $\ess{=}7\kt$) assembly around non-cognate RNA frequently leads to the nucleation of multiple partial capsids on the same RNA; typically these intermediates have incompatible geometries and either fail to combine or form malformed capsids. In the cognate RNA simulations, assembly rapidly nucleates around the HA PS; the LA PSs then enhance growth rates such that assembly is completed before additional partial capsids can nucleate elsewhere on the RNA. The frequency of multiple nucleation events as well as the structural heterogeneity of assembly intermediates are shown in SI Fig.~\ref{fig:highHigh}.

{ {\bf Specificity.} An estimate of the specificity conferred by PSs can be obtained by comparing the assembly dynamics in the presence and absence of PSs. We calculated the probability that, for a given $\ess$ and $\Csalt$, assembly of a well-formed capsid occurs around the cognate RNA before the non-cognate RNA, normalized by the probability of complete assembly around either substrate:
\begin{align}
P_1 = \dfrac{\int_{0}^{\tend} dt P_\text{c}(t)[1-f_\text{nc}(t)]^{r_\text{ex}}}{f_\text{c}(\tend)+(1-[1-f_\text{nc}(\tend)]^{r_\text{ex}})-f_\text{c}(\tend)(1-[1-f_\text{nc}(\tend)]^{r_\text{ex}})}
\label{eq:P1}
\end{align}
where $f_\text{nc}(t)$ and $f_\text{c}(t)$ are the time-dependent yields around non-cognate and cognate RNAs (measured from the simulations whose final yields are shown in Figs.~\ref{relativeYield}D,E), and $P_\text{c}(t)=\frac{d f_\text{c}(t)}{d t}$ is the assembly time probability distribution function for cognate RNAs.
The parameter $r_\text{ex}=c_\text{nc}/c_\text{c}$ is the ratio of non-cognate to cognate RNAs.  Eq.~\ref{eq:P1}  for $r_\text{ex}{=}1$ is shown in Fig.~\ref{competitionApproximation}A.

In a fairly wide range of parameter space, the assembly is 100\% specific for assembly around the cognate RNA. However, at parameters which are optimal for assembly around non-cognate RNA (i.e. where assembly without PSs leads to high-yield, $\ess\in[4-6]\kt$, $\Csalt\in[100-300]$mM), there is essentially no selectivity.  This result highlights the importance of the solution conditions when assessing the role of PSs \textit{in vitro}, and may suggest an explanation for the varying levels of specificity for cognate RNAs observed in \textit{in vitro} experiments (see the Introduction).

We next compare this competition estimate approach to explicit competition simulations which contain a cognate RNA, $\rstoich$ non-cognate RNAs, and 60 pentamer subunits. In these simulations, we define specificity as the fraction of simulations in which the first assembled capsid forms around a cognate RNA. Figure~\ref{competitionApproximation}B presents results for $\rstoich=1$ at several subunit-subunit interaction strengths and salt concentrations. For several parameter sets in that figure, assembly is not productive without PSs, and so as expected selectivity is 100\%.  At the other three parameter sets, which result in incomplete selectivity, the predicted and measured values agree to within error. While \textit{in vitro} assays have typically focused on competition between equal concentrations of cognate and non-cognate RNAs, assembly \textit{in vivo} can occur under a large excess of cellular RNAs  \cite{Eigen1991}. The estimated specificity for $r_\text{ex}{=}10$ is shown in Fig.~\ref{competitionApproximation}C.

\begin{figure}[h]
\centering{\includegraphics[width=0.9\columnwidth]{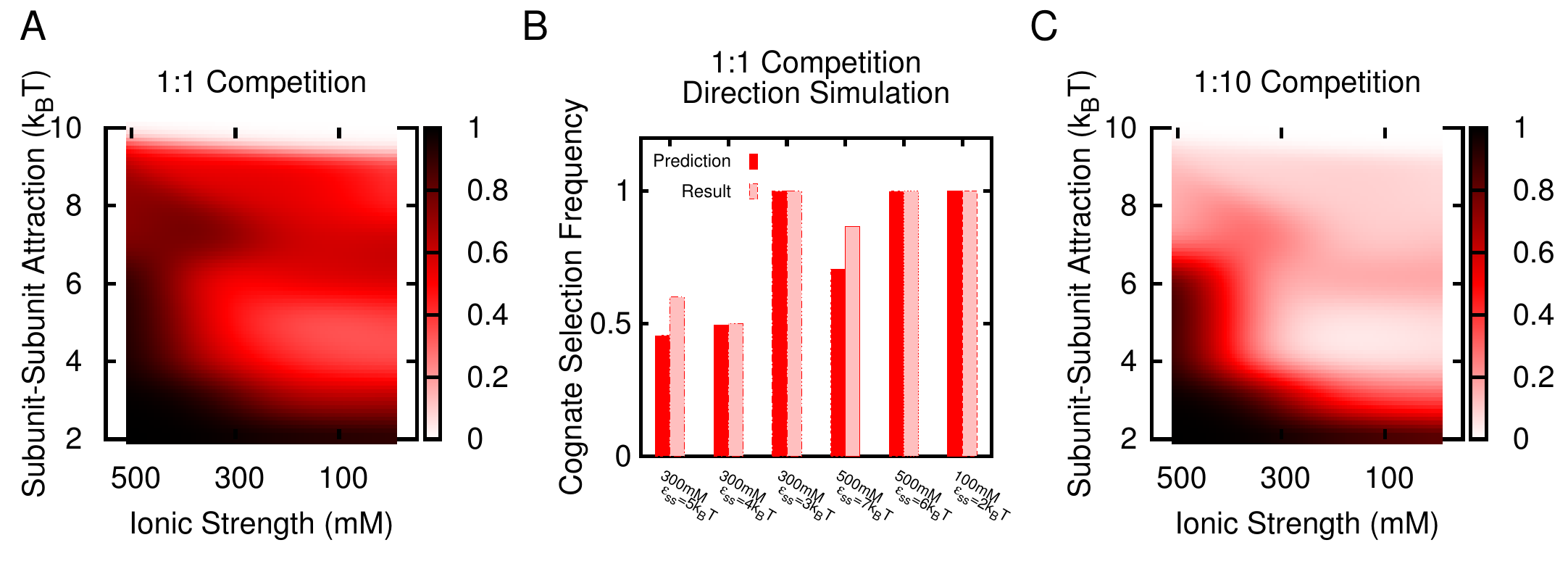}}
\caption{Selectivity for RNA containing 1 HA PS + 25 LA PS competing against a non-cognate RNA at equal concentrations $\rstoich=1$, ({\bf A}) estimated from the data in Fig.~\ref{relativeYield} using Eq.~\ref{eq:P1} and ({\bf B})  measured in direct competition simulations. {\bf (C)} Selectivity for RNA containing 1 HA PS + 25 LA PS competing against excess non-cognate RNA, $\rstoich=10$.   As in Fig~\ref{relativeYield}, in each simulation the optimal RNA length is used based on the results in Fig.~\ref{equilibrium}. In the explicit competition simulations of {\bf (B)} the concentration of subunits is the same as used in the assembly simulations (Fig.~\ref{relativeYield}).}
\label{competitionApproximation}
\end{figure}

We have also considered competition under subunit limiting conditions ---  1 cognate RNA, $\rstoich$ non-cognate RNAs, and 18 protein subunits --- so that at most one complete capsid can assemble, containing either a cognate or a non-cognate RNA. As expected based on the independent assembly simulations, for parameters where assembly is not productive without PS ($\Csalt=500$mM, $\ess=6\kt$), we observe that assembly is 100\% specific for the cognate RNA in all simulations for $\rstoich \in [1-50]$. Interestingly, assembly is also 100\% specific for the cognate RNA  at a parameter set for which assembly is productive without PS ($\Csalt=500$mM, $\ess=7\kt, \rstoich=1$). For these parameters, nucleation occurs first around the PS, which reduces the number of free subunits and impedes nucleation around the non-cognate. We are exploring whether this effect remains in larger systems.

We note that the relationship between $P_1$ and specificity in direct competition assays could break down at low salt, where subunits initially undergo nonspecific absorption onto cognate and non-cognate RNAs. Assembly under limiting subunit concentrations in these conditions requires exchange of subunits between RNAs, which occurs slowly relative to our simulation timescales.

\subsection{The effect of PSs depends on their number and strength}
\label{sec:number}
We now discuss the dependence of assembly and specificity on the number and affinities of PSs. We focus on two interaction parameter sets: $\Csalt{=}100$mM, $\ess{=}2\kt$ and $\Csalt{=}500$ mM, $\ess{=}6\kt$.  These parameters lead to 100\% specificity for the cognate sequence considered in Fig.~\ref{relativeYield} (1 high affinity PS and $\nps{=}25$ low affinity PSs), but represent very different strengths of nonspecific electrostatic interactions and correspondingly different assembly pathways around non-cognate RNAs. For each of these interaction parameter sets, we simulated three distributions of PS affinities along the model RNA (see section~\ref{sec:potentials}):  (\textit{i}) $\nps$ high affinity (HA) PSs, (\textit{ii}) $\nps$ low affinity (LA) PSs, and (\textit{iii}) a `Combo' distribution with 1 HA PS and $\nps$ LA PSs. Recall that the Combo sequence with $\nps{=}25$ is considered in Figs.~\ref{relativeYield} and \ref{competitionApproximation}.

As shown in Fig.~\ref{NumberPS}, assembly yields for both interaction parameter sets are most robust under the Combo PS distribution.  High yields are obtained for intermediate values of $\nps$, although the yield is optimal for sub-stoichiometric $\nps<20$ at moderate salt and super-stoichiometric $\nps>20$ at high salt (recall that there are $\nps=20$ PS binding sites in a complete capsid).

However, the effect of PSs on assembly mechanisms, and hence the dependence on PS distribution, is markedly different for the two parameter sets. At $\Csalt{=}100$ mM, the subunits rapidly adsorb onto the RNA. Without PSs, the weak subunit-subunit interactions ($\ess{=}2\kt$) are insufficient to drive subsequent assembly resulting in a disordered aggregate (Fig.~\ref{relativeYield}A). Nonetheless, even a sub-stoichiometric number of PSs is sufficient to promote complete assembly (Fig.~\ref{NumberPS}A).  High yields are observed for 6-8 HA PSs, 10 LA PSs, and for $\nps \in [10,20]$ for the Combo case. For larger than optimal $\nps$, multiple partial capsids nucleate on the same RNA, leading to long-lived malformed assemblies that suppress yields. Snapshots illustrating typical assembly outcomes at low, stoichiometric, and excess $\nps$ are shown below the plots for each salt concentration in Fig.~\ref{NumberPS}. For the Combo PS sequences, after several subunits have adsorbed onto the RNA, the strong PS initiates assembly, with further growth mediated by the weak PS. When there are multiple HA PSs, it is more likely for multiple small clusters to form, which may then merge into a single capsid, with the final additions driven by electrostatic interactions.

We note that the effect of PSs under low salt and low $\ess$ derives not from their ability to drive subunit-RNA interactions, which are already strong due to nonspecific electrostatics, but rather because the locations of packaging site receptors (PSRs) at the capsid three-fold axes promotes subunit-subunit interactions (Fig.~\ref{schematic}B).  In support of this conclusion, simulations in which PSRs were moved to the center of model subunits led to poor assembly (gold diamond symbol in Fig.~\ref{NumberPS}A).

At high salt concentration ($\Csalt{=}500$ mM), different PS sequences promote assembly (Fig.~\ref{NumberPS}B). Without PSs under these conditions few subunits absorb on the RNA and nucleation does not occur on the timescales being simulated (using Markov State modeling we determined that assembly eventually occurs around the non-cognate RNA on a timescale which is two orders of magnitude longer \cite{Perlmutter2014}). With sub-stoichiometric $\nps$, a cluster of subunits assembles in the vicinity of PSs, but subsequent growth into a capsid is slow on simulated timescales. In contrast, moderate to high yields are observed for super-stoichiometric PSs ($\nps \in [30-40]$). For the Combo sequence, the HA PS promotes rapid nucleation of a trimer after which the LA PSs facilitate adsorption and binding to the cluster by additional subunits.  In contrast to the low salt conditions, super-stoichiometric HA PSs also lead to moderate yields of well-formed capsids; because of the weak nonspecific electrostatics malformed capsids are less prevalent.

With high salt and relatively strong subunit-subunit interactions ($\ess{=}6\kt$) the ability of PSs to drive subunit-RNA interactions should be most relevant to promoting assembly. The effective subunit-subunit interactions promoted by $\nps{=}25$ PSs are stronger than optimal, as can be seen by the fact that reducing $\ess$ increases yields (Fig.~\ref{relativeYield}B). Consistent with this reasoning, eliminating the contribution of PSs to subunit-subunit interactions by moving the PSRs to subunit centers increased the yield (gold diamond symbol in Fig.~\ref{NumberPS}B).

While the location of PSRs within the capsid structure can significantly affect assembly, we found that the location of the strong PS along the RNA (in the Combo sequence) did not measurably alter the yield (open pentagon symbols in Figs.~\ref{NumberPS}A,B). This observation approximately agrees with Dykeman et al. \cite{Dykeman2013a} who predicted a very weak dependence on HA PS location.

\begin{figure}[h]
\centering{\includegraphics[width=0.5\columnwidth]{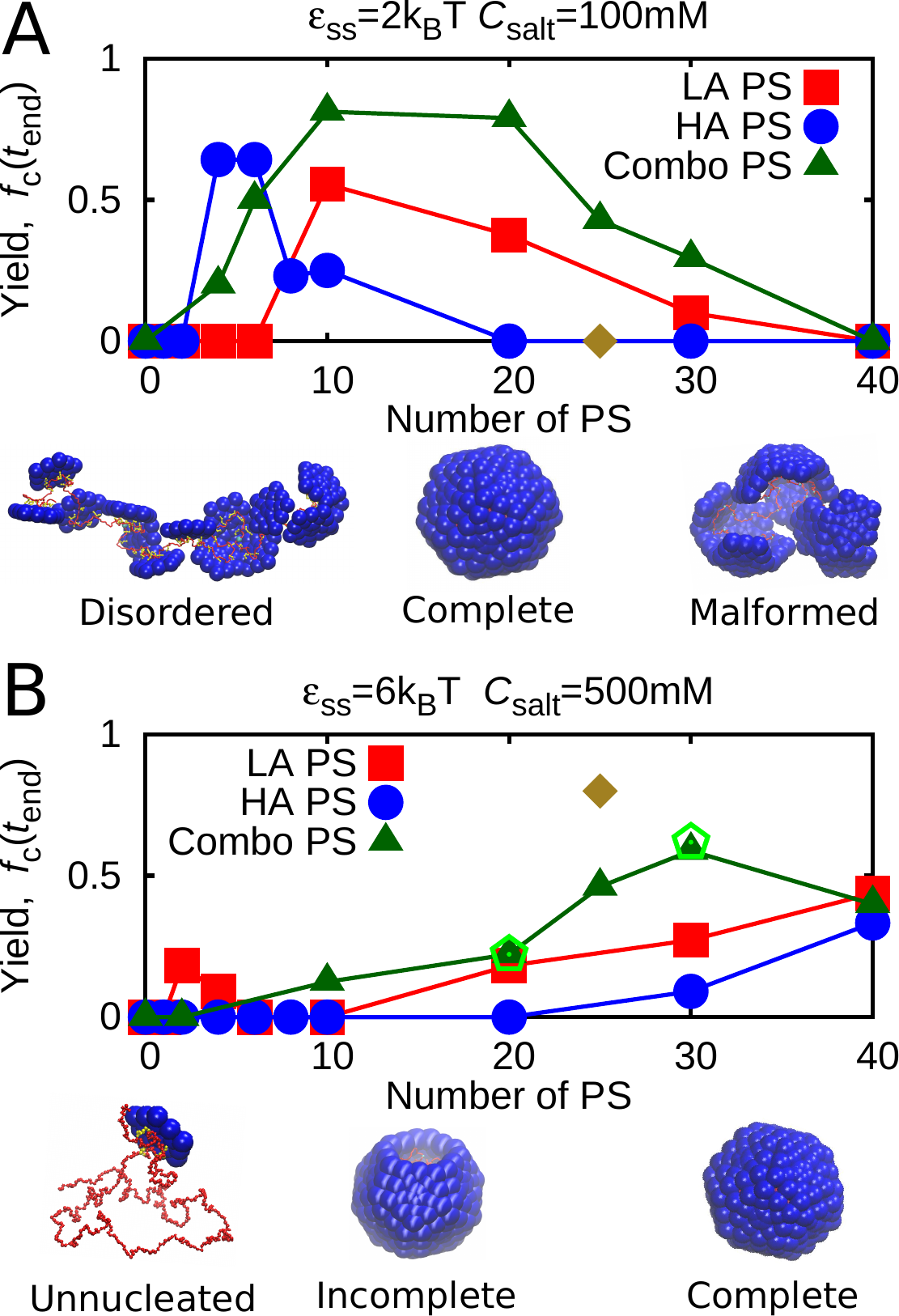}}
\caption{Yield as a function of number of PS, $\nps$, at low {\bf (A)} and high {\bf (B)} salt concentrations. Note that for these parameters yield is zero in the absence of PSs. PSs are either all LA (\textcolor{red}{$\blacksquare$}), all high affinity ({\large\textcolor{blue}{$\bullet$}} symbols), or the Combo sequence with 1 HA and $\nps$ LA PSs(\textcolor{OliveGreen}{$\blacktriangle$} symbols). For these cases, the HA PS is placed in the center of the RNA. Results from sets of simulations with the HA PS placed in the terminal position are shown as {\bf \textcolor{green}{\pentagon}} symbols. The result from simulations with the PS binding site placed in the center of the subunits is shown as a \textcolor{gold(metallic)}{$\blacklozenge$} symbol. Note that there are 20 PS binding sites in a complete capsid, so $\nps=20$ is the stoichiometric value. Snapshots illustrate the trend in dominant outcomes with increasing PS number.}
\label{NumberPS}
\end{figure}

\subsection{PSs can alter assembly pathways}
\label{sec:pathways}
Modeling \cite{Perlmutter2014,Elrad2008,McPherson2005,Hagan2008,Devkota2009,
Hagan2009} and experiments \cite{Kler2012,Cadena-Nava2012,Garmann2014} have shown that assembly pathways around non-cognate RNAs can be classified according to two extremes. Systems in which protein-protein interactions dominate assemble through nucleation-and growth pathways with ordered intermediates, whereas strong protein-RNA interactions (low salt and/or high ARM charges) lead to the `en masse' mechanism in which subunits rapidly adsorb on the RNA in a disordered manner, followed by cooperative rearrangements to form a capsid. Assembly pathways can be classified by the parameter $\nfree$, defined as the number of subunits adsorbed to the RNA which are not part of the largest subunit cluster, averaged over system configurations for which the largest partial capsid intermediate has 4--6 subunits \cite{Perlmutter2014}.  For our model capsid with 12 subunits,  $\nfree \gtrsim 5$ indicates the en masse mechanism, with smaller values indicating the nucleation-and-growth mechanism.  As shown in Fig.~\ref{relativeOrder}, assembly pathways around our model non-cognate RNA range the gamut of $\nfree$, with low salt and low $\ess$ leading to en masse pathways and high salt and high $\ess$ leading to nucleation and growth pathways.

\begin{figure}[h]
\centering{\includegraphics[width=0.75\columnwidth]{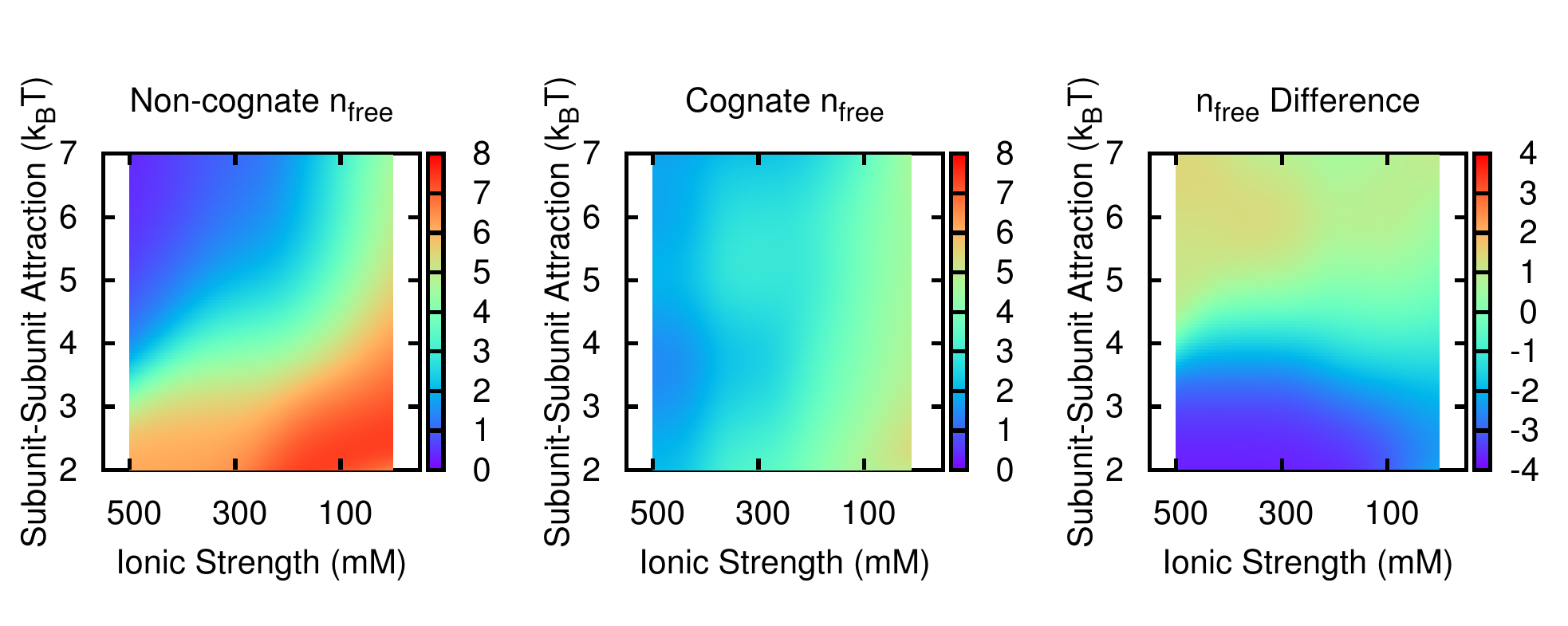}}
\caption{The assembly pathway order parameter $\nfree$ measured from simulations for (left) the non-cognate RNA and (center) the cognate RNA with the Combo PS sequence, $\nps=25$.  (Right) The change in $\nfree$ due to PSs. }
\label{relativeOrder}
\end{figure}

The addition of PSs has a striking effect on assembly pathways.  As shown in Fig.~\ref{relativeOrder}, assembly pathways around the Combo PS sequence with $\nps=25$ correspond to the nucleation-and-growth mechanism over a broad range of $\ess$ and $\Csalt$. Under most conditions the PSs increase the order of assembly intermediates (lowering $\nfree$, Fig.~\ref{relativeOrder}C) because adsorbed subunits are co-localized and well-positioned for assembly.  However, under high $\ess$ the most significant effect of PSs is to increase  subunit adsorption on the RNA and thus PSs slightly increase $\nfree$. Snapshots from representative trajectories for both of these cases are shown in Fig.~\ref{snaps}. Consequently, $\nfree$ and correspondingly the nature of assembly pathways are less sensitive to conditions ($\ess$ and $\Csalt$) than for the non-cognate RNA.  This result parallels the observation that PSs reduce the sensitivity of assembly yields to control parameters (Fig.~\ref{relativeYield}).

\begin{figure}[h]
\centering{\includegraphics[width=0.75\columnwidth]{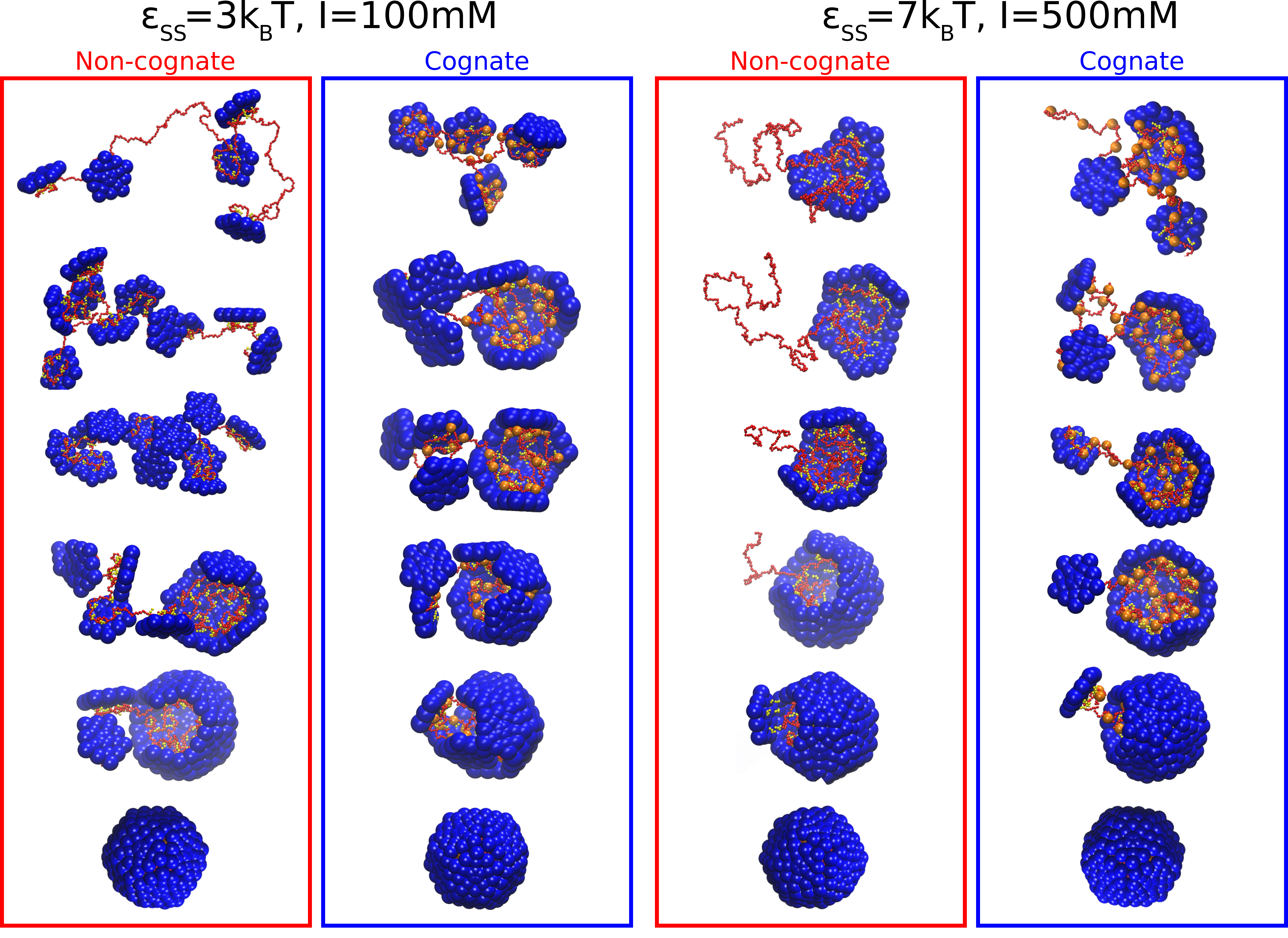}}
\caption{Snapshots from typical assembly trajectories without and with PSs (the cognate here is the combo sequence with 1 HA and 25 LA PSs) for low and high salt concentrations. PSs are depicted as large orange spheres.}
\label{snaps}
\end{figure}

{\bf Relationship between predicted assembly pathways and  single molecule fluorescence correlation spectroscopy (smFCS) data.} A means to test the predicted dependence of assembly pathways on solution conditions and PSs is provided by the fact that pathways with different values of $\nfree$ can be distinguished by the hydrodynamic radii ($\rh$) of their early intermediates \cite{Perlmutter2014}. Recent experiments have used smFCS to monitor the timecourses of $\rh$ during assembly around cognate and non-cognate RNAs \cite{Borodavka2012,Comas-Garcia2014, Patel2015}. Under the experimental conditions, assembly around cognate RNAs was rapid and characterized by either constant $\rh$ or a collapsed complex followed by gradual increase to the size of an assembled capsid. Assembly around non-cognate RNAs was slower, with $\rh$ initially increasing before finally decreasing to the size of the capsid.

To relate the predicted effect of PS on assembly pathways ($\nfree$) discussed above to an experimentally observable quantity,
we estimated the hydrodynamic radii $\rh$ for polymer-subunit intermediates using the program HYDROPRO, which has been shown to accurately predict $\rh$ for large protein and protein-NA complexes \cite{Ortega2011}. Fig.~\ref{radius} shows calculated $\rh$ for assembly around cognate and non-cognate RNAs for $\Csalt{=}100mM$ and several values of $\ess$. For weak subunit-subunit interactions ($\ess=2\kt$, Fig.~\ref{radius}A), assembly without PSs results in disordered aggregates (Fig.~\ref{relativeYield}), and the $\rh$ monotonically increases over time. With PSs, on the other hand, the $\rh$ initially increases as subunits attach to the RNA, and then rapidly decreases as the capsid assembles. Upon increasing the subunit-subunit interactions strength ($\ess=3\kt$, Fig.~\ref{radius}B), successful assembly occurs with and without PS; however, the increase in $\rh$ is greater and of longer duration in the absence of PS. This difference occurs because the PSs enhance the assembly rate and decrease $\nfree$. Finally, under stronger subunit-subunit interactions ($\ess=4\kt$) PSs have little effect on $\nfree$ (Fig.~\ref{relativeOrder}) and correspondingly the time course of $\rh$ (Fig.~\ref{radius}C) is similar for cognate and non-cognate RNAs.

The results at low subunit-subunit interaction strength resemble some key features of the experimental observations of $\rh$ around cognate and non-cognate RNAs, while the lack of effect of PSs on assembly pathways under larger $\ess$ emphasizes the fact that specificity depends on the underlying assembly driving forces. Importantly, the ability to change $\rh$ time courses does not depend on specific geometric features of the PSs in our simulations, but only requires that PSs promote rapid, ordered assembly pathways.  Increasing the subunit-subunit interactions can achieve a similar effect as adding PSs in this regard (Fig.~\ref{radius}C).

In the simulations discussed thus far (Fig.~\ref{relativeYield}A-C), we used a relatively short RNA (575 segments at $\Csalt=100mM$), since this is the optimal length for our subunits with the small ARM charge (+5) \cite{Perlmutter2013,Perlmutter2014}.  Therefore, the $\rh$ of the free RNA (prior to encapsidation) is less than that of the assembled capsid and the measured $\rh$ increases substantially upon adsorption of subunits, until assembly of ordered partial capsids reduces $\rh$.  To examine the applicability of our findings to the more typical case in which the free RNA $\rh$ is similar to or larger than the capsid size, we also performed simulations on subunits with ARMs with charge (+10) and optimal RNA length 910 segments (Fig.~\ref{radius}D-F). The behavior of this system is qualitatively similar to that of the +5 ARMs, except that the initial increase in $\rh$ upon subunit absorption is less apparent for ordered assembly pathways. Note that in previous simulations \cite{Perlmutter2014} we found that including base pairing did not qualitatively change $\rh$ time courses.  However, we note that we do not observe an increase in $\rh$ after the initial decrease, as observed in some of the experiments \cite{Borodavka2012,Patel2015}.  This pattern might reflect a global RNA conformation change triggered by subunit binding, which is not currently considered by our model.

\begin{figure}
\centering{\includegraphics[width=0.75\columnwidth]{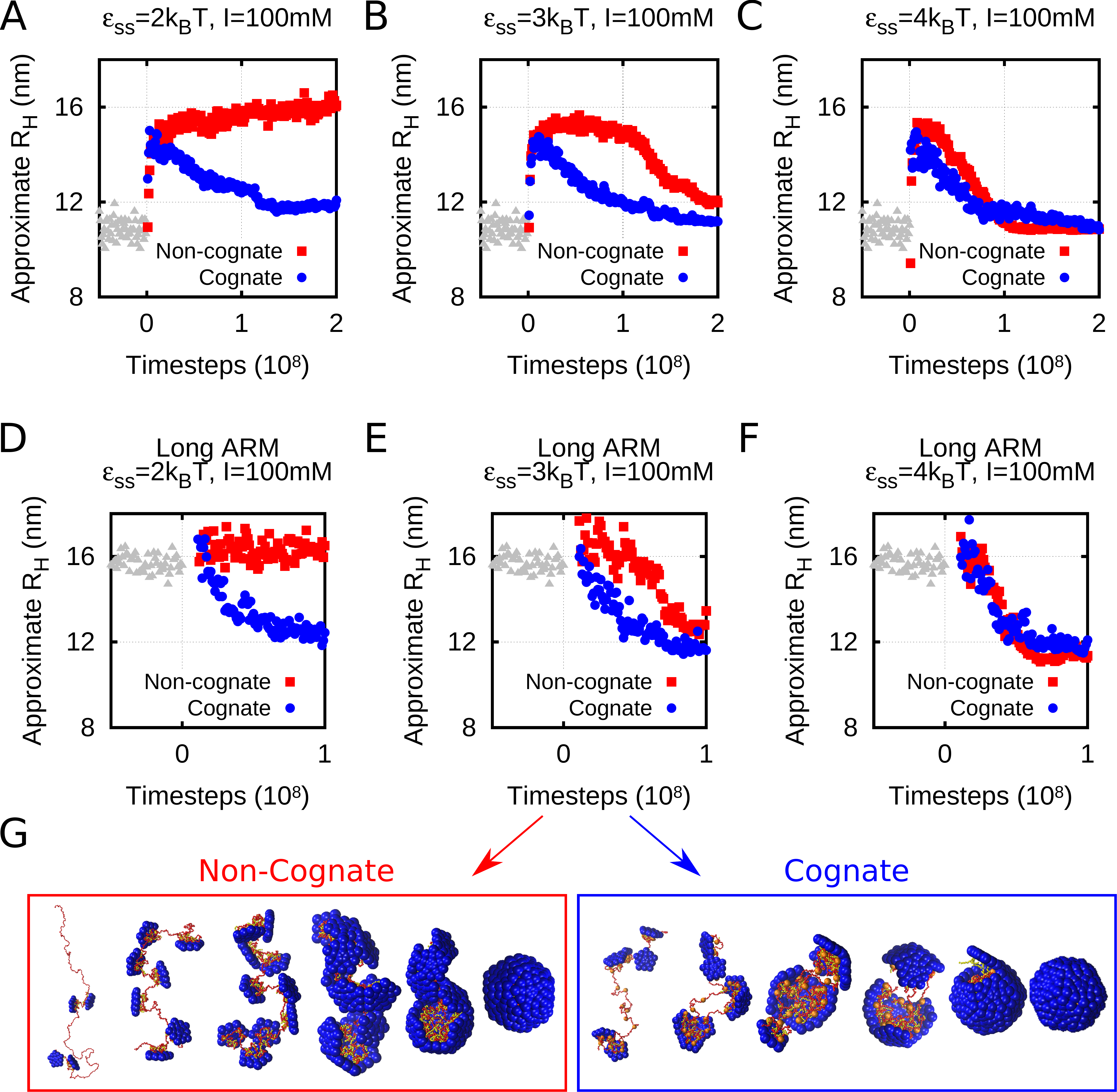}}
\caption{{\bf (A-F)} Radius of hydration $\rh$ as a function of simulation time steps for assembly trajectories performed at indicated parameter values, for non-cognate RNA (\textcolor{red}{$\blacksquare$} symbols) and cognate RNA ({\large \textcolor{blue}{$\bullet$}} symbols). The $\rh$ values before subunits are introduced are shown as \textcolor{gray}{$\blacktriangle$} symbols. The subunit-subunit interaction energy $\ess$ increases from left to right. In the top row (A-C), the subunit ARM charge is (+5), and the RNA length is 575 segments; in the second row (D-F), the subunit ARM charge is (+10), and the RNA length is 910 segments. {\bf (G)} Snapshots from simulations corresponding to panel (E), with non-cognate RNA on the left and cognate RNA on the right. }
\label{radius}
\end{figure}

\subsection{Restricted RNA Conformational Dynamics Inhibit Assembly}

Although our simulations show that PSs can promote efficient assembly under conditions in which nonspecific assembly fails, we also found that non-optimal numbers and affinities of PSs can lead to kinetic traps.  In this section we discuss mechanisms that can lead to these kinetic traps. As seen in Fig.~\ref{NumberPS}, RNAs with 20 HA PSs (the stoichiometric amount) fails to produce capsids at high or low salt. In part this outcome follows the well-known rule that strong interactions hinder self-assembly into highly ordered low free energy configurations by preventing `local' equilibration among partially assembled configurations and thus trapping the system in metastable disordered states \cite{Ceres2002, Zlotnick2003,Hagan2006, Jack2007, Rapaport2008, Hagan2011, Grant2011, Hagan2014, Whitelam2014}, as found for strong nonspecific electrostatics ($\Csalt \leq10$ mM) or subunit-subunit interactions ($\ess\geq8\kt$, Fig.~\ref{schematic}). However, a more nuanced explanation is required, since at high salt adding more HA PSs ($\nps=[30,40], \eps=20\kt$) does lead to moderate assembly yields (Fig.~\ref{NumberPS}).  Analysis of the high salt, $\nps=20$ simulation trajectories revealed ordered partially assembled capsid intermediates which, despite having optimal subunit-subunit interaction geometries, failed to reach completion (Fig.~\ref{pathExploration}).

We hypothesized that the stalled assemblies result from poorly equilibrated RNA conformational dynamics within assembling capsids.  To investigate the dependence of the RNA dynamics on PS sequences, we measured the fluctuations of the paths traced by RNAs within capsids (or capsid intermediates) over the course of dynamical trajectories. To simplify the analysis we define an RNA path as the sequence of capsid threefold sites with which RNA segments strongly interact (counting by the index of each strongly interacting RNA segment, Fig.~\ref{pathExploration}B).  Because the ARM anchoring sites and the PSRs are located near threefold sites, these sites have enhanced non-cognate RNA densities both in the presence and absence of PSs \cite{Perlmutter2013}. We note that these paths can be complex; for example, with strong electrostatics ($\Csalt=100mM$), $\sim50\%$ of paths re-visit one or more threefold sites (meaning segments which are nonlocal in sequence interact with the same threefold site), compared with only $\sim10\%$ at ($\Csalt=500mM$). Similarly, jumps between non-neighboring vertices are also common, present in $\sim1/3$ of paths. Thus, the RNA usually does not trace a Hamiltonian path within the capsid in our simulations.

In Figure~\ref{pathExploration} we quantify the RNA path persistence time (time required to change conformation) within assembled capsids. In Fig.~\ref{pathExploration}A, we see that non-cognate RNAs and RNAs with the Combo PS sequences (primarily LA PSs) are highly dynamic; a new RNA path is observed at almost every observation time. In contrast, an RNA with 20 HA PSs is far less dynamic, with existing paths persisting for far longer periods of time.  To help visualize the differences between these two classes of dynamics, schematics of pathways observed at five different observation times for different RNA sequences are shown in Fig.~\ref{pathExploration}B.  While the paths for the 20 HA PS sequence are nearly identical, the paths change significantly on this timescale for the other two PS sequences. We find that assembly stalls when the RNA becomes frozen in a conformation whose geometry hinders recruitment of additional subunits to the assembling capsid. Two examples of such conformations are shown in Fig.~\ref{pathExploration}C. Under parameters which promote RNA dynamics, such conformations are transient.

Notably, as the number of PSs is increased beyond the stoichiometric value $\nps=20$ for any distribution, the RNA dynamics increases (Fig.~\ref{polDynamics}).  This observation explains the increase in yield for the HA PS distribution at high $\nps$.
This facilitation arises because excess PSs can displace bound PSs from a given three-fold site without requiring complete loss of interaction at that site, thus avoiding the activation barrier associated with PS-PSR unbinding (see SI section \ref{sec:Stalling} and Fig.~\ref{polDynamics}).  In contrast, for $\nps \le 20$ exchange requires dissolution of a PS-PSR interaction. In effect, RNA with excess PSs can slide on capsid intermediates to sample different conformations and thus escape from unproductive traps.
Previous simulations \cite{Kivenson2010,Elrad2010} and a theoretical model \cite{Hu2007} have suggested the importance of RNA rearrangements and subunit `sliding' during assembly around non-cognate polyelectrolytes .

\begin{figure}[h]
\centering{\includegraphics[width=0.75\columnwidth]{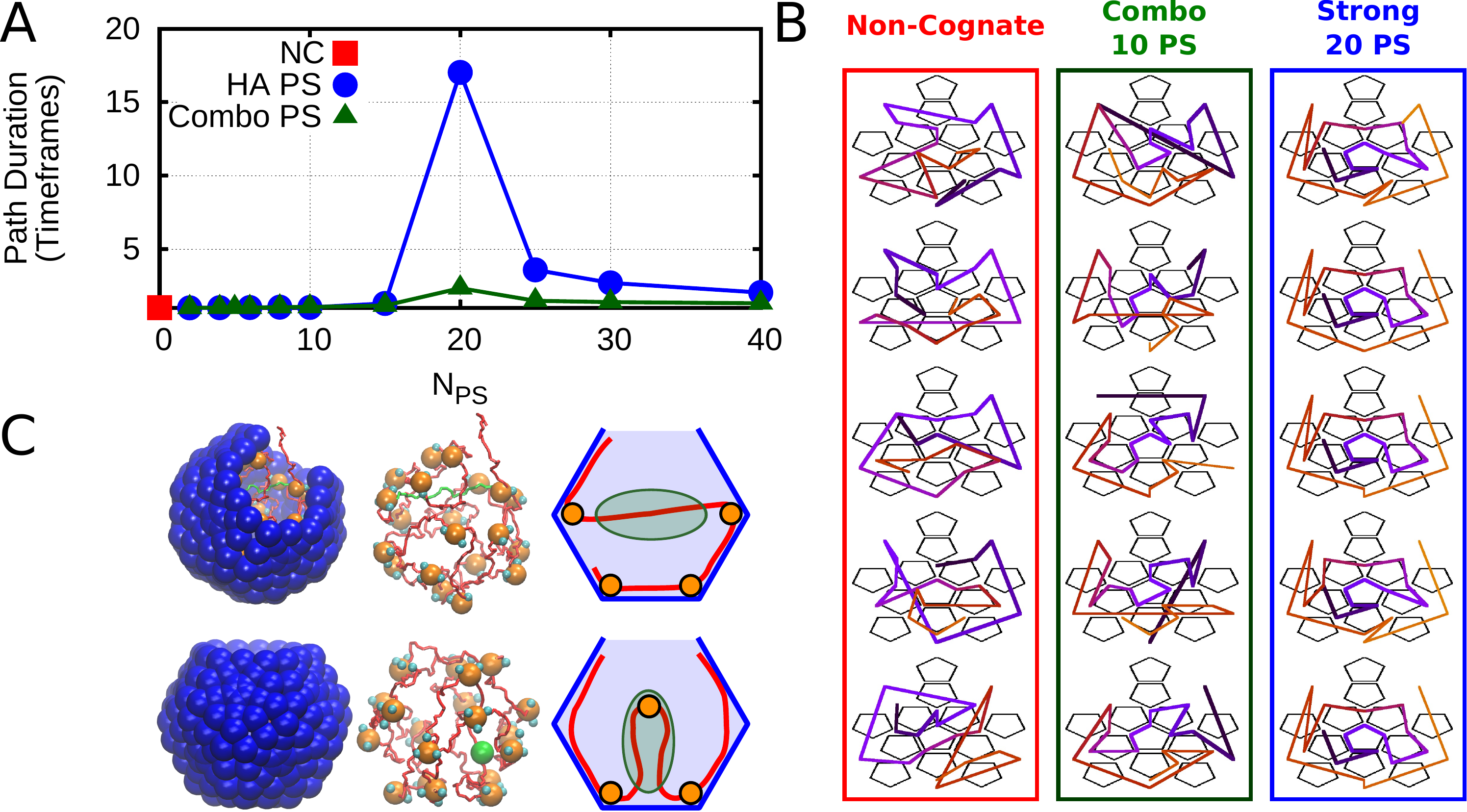}}
\caption{\textbf{(A)} Rate of path discovery during a dynamic trajectory for RNAs within a preassembled capsid. \textbf{(B)} Schematic representation of RNA path within the capsid at intervals of $5\times10^{7}$ timesteps. Line indicates path of RNA, with line color and width changing gradually with contour length for clarity. \textbf{(C)} Snapshots and schematics indicating non-optimal RNA paths which lead to stalled assemblies. PSs are shown with exaggerated size to improve visibility. Segments of interest are shown in green.}
\label{pathExploration}
\end{figure}

\section{Discussion}
In this article, we have described simulations of capsid assembly around RNA, represented as a flexible polyelectrolyte with sequence-specific protein-RNA interactions, or packaging sites (PSs).  By performing extensive simulations over a range of ionic strengths, simulated protein-protein interaction strengths, and strength and number of PSs, we have explored how PSs alter the pathways and products of capsid assembly reactions, and the extent to which they induce specificity against polyelectrolytes without sequence-specific interactions (e.g. non-cognate RNAs). We find that PSs can confer arbitrarily high specificity over RNAs with uniform nonspecific interactions, but that the degree of specificity is sensitive to the underlying assembly driving forces, which can be tuned by solution conditions (ionic strength, $p$H) as well as capsid protein charge (ARM sequence).  The best specificity is conferred under conditions where the nonspecific interactions alone are slightly too weak to promote effective assembly.

Assembly and specificity are also sensitive to the affinity and number of PSs, with the optimal distribution of PSs depending on the solution conditions.  Our simulations suggest that the PS sequences that confer the highest specificity and are most robust to solution conditions contain one or a few high affinity PSs and a stoichiometric or small excess of low affinity PSs.  This observation is consistent with recent models \cite{Dykeman2013a}, observation of multiple weak PSs in viral genomes \cite{Dykeman2013b}, and recent \textit{in vitro} measurements \cite{Patel2015}. Our simulations identify multiple mechanisms by which PSs can confer specificity, depending on the protein sequence and solution conditions.  Under conditions where protein subunit-subunit or sequence-independent protein-RNA interactions are too weak to nucleate assembly, PSs that enhance protein-RNA interactions and RNA-mediated protein-protein interactions can induce nucleation and facilitate subsequent assembly.  Under conditions where strong subunit-subunit and nonspecific subunit-RNA interactions lead to multiple, geometrically incompatible partial capsids forming on individual RNAs, efficient and specific assembly can be realized by PS sequences  that favor nucleation and rapid assembly of a single partial capsid.  Finally, the simulations demonstrate that PSs can dramatically alter assembly pathways in comparison to non-cognate RNAs (Fig.~\ref{radius}), as observed in recent experiments \cite{Borodavka2012}, but that the effect on assembly pathways is sensitive to solution conditions.

Our simulation results suggest that rapid, specific assembly can proceed via a diverse ensemble of pathways, provided that RNA conformations can anneal during assembly through reversible interactions and/or cooperative RNA-protein rearrangements. This finding is consistent with the observation that proteins can fold by multiple, dissimilar pathways \cite{Bowman2010a}.
In particular, the RNA does not trace a Hamiltonian path, as has been inferred from structures of $T{=}3$ MS2 capsids \cite{Dykeman2011} and assumed in other models \cite{Dykeman2013a,Dykeman2014}. However, the expectation of a Hamiltonian path relies on coupling between PS binding and subunit conformation, which is not present in our model of a $T{=}1$ capsid.

Dykeman et al. \cite{Dykeman2014} recently showed that the gradually increasing protein concentration characteristic of an MS2-infected E. coli can increase specificity for a cognate RNA in comparison to assembly under a fixed protein concentration.  Enhanced specificity arises in their model because during the initial stages of the reaction nucleation occurs only around cognate RNAs, similar to the behavior in our simulations for high salt and low $\ess$ (Fig.~\ref{competitionApproximation}).  When the protein concentration increases during later stages it is rapidly consumed by growth of the partial capsid-cognate RNA complexes.  While simulations with time-varying protein concentrations are beyond the scope of the present work, we anticipate that similar specificity enhancements would arise in our model.

{\bf Implications for experiments.}
Our simulations predict that the specificity conferred by PSs is sensitive to parameters that control the pathways and efficiency of sequence-independent assembly. While it has been previously suggested that the degree of specificity observed in \invitro experiments is sensitive to subunit concentration, the predicted phase diagrams reveal that varying ionic strength, $p$H, or protein-RNA binding sequences (through mutagenesis) could shift an experiment from selective to nonselective.  This result may shed light on the varied degrees of specificity observed in previous competition experiments \cite{Ling1970,Sorger1986,Beckett1988,Porterfield2010,Comas-Garcia2012}.  However, note that the location of boundaries within the predicted phase diagrams depend on the protein-RNA binding sequence\cite{Perlmutter2014}. Furthermore, experimentally measured specificity has been defined in different ways, depending on whether capsid protein is in excess or limiting. Our simulations find that these two conditions and definitions can lead to similar or different observed levels of specificity, depending on the location and parameter space (see section ~\ref{sec:specificity}).
An intriguing prediction from our simulations is that excess PSs (in comparison to the number of binding sites within a complete capsid) can increase assembly under some conditions by promoting exchange of improperly bound PSs.  This prediction, as well as the general dependencies on the affinity and number of PSs, could be tested by constructing RNA fragments with varying numbers of high- and low-affinity PSs.

\section{Acknowledgments}
We gratefully acknowledge William Gelbart and Chuck Knobler for insightful discussions and critical reading of the manuscript, as well as Reidun Twarock for a critical reading of the manuscript and helpful discussions, particularly about the geometry of packaging site--capsid protein interactions. This work was supported by Award Number R01GM108021 from the National Institute Of General Medical Sciences. Computational resources were provided by the NSF through XSEDE computing resources (Maverick and Keeneland) and the Brandeis HPCC which is partially supported by the Brandeis Center for Bioinspired Soft Materials, an NSF MRSEC,  DMR-1420382.

\bibliographystyle{unsrt}
\bibliography{all-references}

\pagebreak
\pagebreak
\setcounter{figure}{0}
\setcounter{section}{0}
\renewcommand{\thefigure}{S\arabic{figure}}
\renewcommand{\thesection}{S\Roman{section}}

{\large{\bf Supporting Information} \\}

\section{Model and simulation details}
\subsection{Model details}
\label{sec:potentials}
In our model, all potentials can be decomposed into pairwise interactions. Potentials involving capsid subunits further decompose into pairwise interactions between their constituent building blocks -- the excluders, attractors, `Top' and `Bottom', and ARM pseudoatoms. It is convenient to state total energy of the system as the sum of 6 terms: a capsid subunit - subunit $U\sub{ss}{}$ part (which does not include interactions between ARM pseudoatoms), subunit-ARM $U\sub{sa}{}$, polymer-polymer (i.e. RNA-RNA) $U\sub{pp}{}$, ARM-ARM $U\sub{aa}{}$, polymer-ARM $U\sub{pa}{}$, and subunit-polymer $U\sub{sp}{}$ (which includes the PS-PSR interactions), each summed over all pairs of the appropriate type:
\begin{align}
U = & \sum_{\mathrm{sub\ }{i}} \sum_{\mathrm{sub\ }{j < i}} U\sub{ss}{}
  + \sum_{\mathrm{sub\ }{i}} \sum_{\mathrm{ARM\ }{j}} U\sub{sa}{}
  + \sum_{\mathrm{poly\ }{i}} \sum_{\mathrm{poly\ }{j < i}} U\sub{pp}{}
  + \sum_{\mathrm{ARM\ }{i}} \sum_{\mathrm{ARM\ }{j < i}} U\sub{aa}{} \nonumber \\
  & + \sum_{\mathrm{poly\ }{i}} \sum_{\mathrm{ARM\ }{j}} U\sub{pa}{}
  + \sum_{\mathrm{sub\ }{i}} \sum_{\mathrm{poly\ }{j}} U\sub{sp}{}
\end{align}
where $\sum_{\mathrm{sub\ }{i}} \sum_{\mathrm{sub\ }{j < i}}$ is the sum over all distinct pairs of capsid subunits in the system, $\sum_{\mathrm{sub\ }{i}} \sum_{\mathrm{poly\ }{j}}$ is the sum over all subunit-polymer pairs, etc. Note that unless otherwise stated, polyelectrolyte segments and polymer segments designated as PS have the same interactions and parameters. This rule is excepted for PS-PS interactions and PS-PSR interactions, as described below.

The capsid subunit-subunit potential $U\sub{ss}{}$ is the sum of the attractive interactions between complementary attractors, and geometry guiding repulsive interactions between `Top' - `Top' pairs and `Top' - `Bottom' pairs. There are no interactions between members of the same rigid body, but ARMs are not rigid and thus there are intra-subunit ARM-ARM interactions. Thus, for notational clarity, we index rigid bodies and non-rigid pseudoatoms in Roman, while the pseudoatoms comprising a particular rigid body are indexed in Greek. For subunit $i$ we denote its  attractor positions as $\{\mathbf{a}_{i\alpha}\}$ with the set comprising all attractors $\alpha$, its `Top' positions $\{\mathbf t_{i\alpha}\}$, and its `Bottom' positions $\{\mathbf b_{i\alpha}\}$. The capsid subunit-subunit interaction potential between two subunits $i$ and $j$ is then defined as:

\begin{eqnarray}
\label{Uss}
U\sub{cc}{}(\{\mathbf a_{i\alpha}\}, \{\mathbf t_{i\alpha}\}, \{\mathbf b_{i\alpha}\} , \{\mathbf a_{j\beta}\}, \{\mathbf t_{j\beta}\}, \{\mathbf b_{j\beta}\})  &=&
    \sum_{\alpha,\beta}^{N\sub{t}{}} \varepsilon \LJ{} \left(
    \left|\mathbf{t}_{i\alpha} - \mathbf{t}_{j\beta} \right|,
    \ \sigma\sub{t}{} \right)
    \nonumber \\
    &+&
	\sum_{\alpha,\beta}^{N\sub{b}{},N\sub{t}{}} \varepsilon \LJ{} \left(
    \left| \mathbf{b}_{i\alpha} - \mathbf{t}_{j\beta} \right|,
    \ \sigma\sub{b}{} \right)
    \nonumber \\
    &+&
    \sum_{\alpha,\beta}^{N\sub{a}{}} \varepsilon \Morse{} \left(
    \left|\mathbf{a}_{i\alpha} - \mathbf{a}_{j\beta} \right|,
    \ r\sub{0}{}, \varrho, \rcut \right)
    \nonumber \\
\end{eqnarray}
where $\varepsilon$ is an adjustable parameter which both sets the strength of the capsid subunit-subunit attraction at each attractor site and scales the repulsive interactions which enforce the dodecahedral geometry, $N\sub{t}{}$, $N\sub{b}{}$, and $N\sub{a}{}$ are the number of `Top', `Bottom', and attractors pseudoatoms respectively in one subunit, $\sigma\sub{t}{}$ and $\sigma\sub{b}{}$ are the effective diameters of the `Top' - `Top' interaction and `Bottom' - `Top' interaction, which are set to $10.5$ nm and $9$ nm, respectively. $r\sub{0}{}$ is the minimum energy attractor distance, set to $1$ nm, $\varrho$ is a parameter determining the width of the attractive interaction, set to $2.5$, and $\rcut$ is the cutoff distance for the attractor potential set to $10$ nm.

The function $\LJ{}$ is defined as the repulsive component of the Lennard-Jones potential shifted to zero at the interaction diameter:
\begin{equation}
\LJ{}(x,\sigma) \equiv
\left\{  \begin{array}{ll}
	\left(\frac{\sigma}{x}\right)^{12} -1 & : x < \sigma \\
	0 & : \mathrm{otherwise}
\end{array} \right.
\label{eq:LJ}
\end{equation}
The function $\Morse{}$ is a Morse potential:
\begin{equation}
\Morse{}(x,r\sub{0}{},\varrho) \equiv
\left\{  \begin{array}{ll}
\left(e^{\varrho\left(1-\frac{x}{r\sub{0}{}}\right)} - 2 \right)e^{\varrho\left(1-\frac{x}{r\sub{0}{}}\right)} - V_\text{shift}(\rcut) & : x < \rcut \\
 0 & : \mathrm{otherwise}
\end{array} \right.
\label{eq:Morse}
\end{equation}
with $V_\text{shift}(\rcut)$ the value of the unshifted potential at $\rcut$.

The capsid subunit-ARM interaction is composed of a short-range repulsion representing the excluded volume. For subunit $i$ with excluder positions $\{\mathbf x_{i\alpha}\}$ and ARM segment $j$ with position $\mathbf R_j$, the potential is:
\begin{eqnarray}
\label{Usa}
U\sub{sa}{}(\{\mathbf x_{i\alpha}\}, \mathbf R_j) &=&
    \sum_{\alpha}^{N\sub{x}{}} \LJ{} \left(
    | \mathbf{x}_{i\alpha} - \mathbf R_j |,
    \sigma\sub{xA}{}\right)
\end{eqnarray}
$\sigma\sub{xa}{} = 0.5(\sigma_\text{x}+\sigma_\text{a})$ is the effective diameter of the excluder-ARM repulsion with $\sigma_\text{a}=0.5$ nm the diameter of an ARM bead.

The interactions between polymer segments are defined as follows.
Polymer segments which occupy adjacent positions within a polymer chain experience only a harmonic potential $\mathcal{K}_\mathrm{bond}$ which depends on bond distance. The polymer-polymer non-bonded interaction is composed of electrostatic repulsion and short-ranged excluded volume interactions, as well as an additional longer-range repulsion between PS segments added to prevent multiple PS segments from occupying a single binding site. The potential between two polymer segments $i$ and $j$ is given by
\begin{align}
\label{Upp}
U\sub{pp}{}(\mathbf R_i, \mathbf R_j, \mathbf R_k)
    &=&
    \left\{
            \begin{array}{ll}
                \mathcal{K}_\mathrm{bond}(R_{ij}, \sigma\sub{p}{},k_\mathrm{bond})    & : \{i,j\}\ \mathrm{bonded} \\
                \LJ{}(R_{ij}, \sigma\sub{p}{}) + \mathcal{I}_\text{ps}(i)\mathcal{I}_\text{ps}(j) \LJ{}(R_{ij}, \sigma\sub{ps}{}) + \mathcal{U}\rsub{DH}(R_{ij},\qp,\qp, \sigma\sub{p}{})                        & : \{i,j\}\ \mathrm{nonbonded}\\
    	\end{array}
	\right.
\end{align}
where $R_{ij} \equiv | \mathbf R_i - \mathbf R_j |$ is the center-to-center distance between the polymer segments and   $\sigma\sub{p}{}=0.5$nm is the diameter of a polymer segment.  The harmonic bond pontential
between sequential segments given by
\begin{equation}
    \mathcal{K}_\mathrm{bond}(R_{ij}, \sigma, k_\mathrm{bond})
    \equiv
    	\frac{k_\mathrm{bond}}{2}(R_{ij}-\sigma)^2.
\end{equation}
 The function $\mathcal{I}_\text{ps}(k)=1$ if segment $i$ is a  packaging site and 0 otherwise.  This interaction term accounts for steric interactions which inhibit multiple packaging sites from interacting with the same site on a capsid protein by adding an additional repulsive interaction with effective diameter $\sigma\sub{ps}{}=3$nm.
The final term in Eq.~\ref{Upp} is a Debye-H\"{u}ckel potential accounting for screened electrostatic interactions between polymer segments with valence charge $\qp=-1$, given by
\begin{equation}
\mathcal{U}\rsub{DH}(r,q_1,q_2, \sigma{})/\kt \equiv
\left\{  \begin{array}{ll}
	\frac{q_1 q_2  l\rsub{b} \lambda_\text{D}\ e^{\sigma{}/\lambda\rsub{D}}}{\lambda\rsub{D}+\sigma{}} \left(\frac{e^{-r/\lambda\rsub{D}}}{r} \right) & : r < 2\lambda\rsub{D} \\
	 \frac{(r_{cut}^{2}-r^{2})^{2}(r_{cut}^{2}+2r^{2}-3r_{on}^{2}))}{(r_{cut}^{2}-2r_{on}^{2})^{3}}
	\frac{q_1 q_2  l\rsub{b} \lambda_\text{D}\ e^{\sigma{}/\lambda\rsub{D}}}{\lambda\rsub{D}+\sigma{}}
	\left(\frac{e^{-r/\lambda\rsub{D}}}{r} \right) & : 2\lambda\rsub{D} < r < 3\lambda\rsub{D} \\
	0 & : \mathrm{otherwise}
\end{array} \right.
\end{equation}
 with
$\lambda\rsub{D}$ as the Debye length, $l\rsub{b}$ as the Bjerrum length, $q_1$ and $q_2$ as the valences of the interacting charges, and the potential is smoothly switched to zero at the cutoff distance $r{=}3\lambda\rsub{D}$.

The ARM-ARM interaction is similar to the polymer-polymer interaction, consisting of non-bonded interactions composed of electrostatic repulsions and short-ranged excluded volume interactions, as well as bonded interactions between sequential monomers in the ARM chain:

\begin{eqnarray}
\label{Uaa}
U\sub{aa}{}(\mathbf R_i, \mathbf R_j)
    &=&
    \left\{
            \begin{array}{ll}
                \mathcal{K}_\mathrm{bond}(R_{ij}, \sigma\sub{a}{},k_\mathrm{bond})    & : \{i,j\}\ \mathrm{bonded} \\
                \LJ{}(R_{ij}, \sigma\sub{a}{}) + \mathcal{U}\rsub{DH}(R_{ij},q_i,q_j, \sigma\sub{a}{})                        & : \{i,j\}\ \mathrm{nonbonded}\\
            \end{array}
        \right.
\end{eqnarray}
where $R_{ij} \equiv | \mathbf R_i - \mathbf R_j |$ is the center-to-center distance between the ARM subunits and $q_i$ is the valence of charge on ARM segment $i$. For the simulations described in this work, $q_i = -1$ for all ARM subunits.

The ARM-Polymer interaction is the sum of repulsive, short-ranged excluded volume interactions and electrostatic interactions:
\begin{eqnarray}
\label{Upa}
U\sub{pa}{}(\mathbf R_i, \mathbf R_j)
    &=&
	\LJ{}(R_{ij}, \sigma\sub{ap}{}) + \mathcal{U}\rsub{DH}(R_{ij},q_i,q_j, \sigma\sub{ap}{})
\end{eqnarray}
with $\sigma\sub{ap}{}=0.5$nm.

The capsid subunit-polymer interaction is a short-ranged repulsion representing the excluded volume, with an additional attractive interaction between the packaging site receptor (PSR) on the subunit and polymer segments which are packaging sites (PS). For capsid subunit $i$ with excluder positions $\{\mathbf{x}_{i\alpha}\}$ and PSR $\{\mathbf{c}_{i\alpha}\}$ and polymer segment $j$ with position $\mathbf R_j$, the potential is:
\begin{eqnarray}
\label{Usp}
U\sub{sp}{}(\{\mathbf x_{i\alpha}\}, \mathbf R_j) &=&
    \sum_{\alpha}^{N\sub{x}{}} \LJ{} \left(
    | \mathbf{x}_{i\alpha} - \mathbf R_j |,
    \sigma\sub{xp}{}\right)
	+
    \mathcal{I}_\text{ps}(j)\sum_{\alpha}^{N\sub{c}{}} \eps \Morse{} \left(
    \left|\mathbf{c}_{i\alpha} -  \mathbf R_j \right|,
    \ r\sub{0}{}, \varrho, \rcut \right)
\end{eqnarray}
where $N_\text{x}$ is the number of excluders on a capsid subunit, $N_\text{c}$ is the number of PSRs on a subunit, $\sigma\sub{xp}{} = 0.5(\sigma_\text{x}+\sigma_\text{p})$ is the effective diameter of the excluder-polymer repulsion with $\sigma_\text{x}=3$ nm and $\sigma_\text{p}=0.5$ nm the respective diameters of excluder and polymer beads. The Morse parameters used for the PS-PSR interaction are the same as used in the the attractor-attractor interaction, except for $\eps$ which is set at $5$ or $20\kt$, as discussed in the main text. Note that the second term of equation~\ref{Usp} only applies to the polymer segments ($j$) which are PSs.

\subsection{Simulations}
Trajectories are simulated using the Brownian Dynamics algorithm of HOOMD, which uses the Langevin equation to calculate the time evolution of positions and rigid body orientations \cite{Anderson2008}. For each of our dynamical assembly simulations, the $\mbox{box size} = 200 \times 200 \times 200$ nm. Except where mentioned otherwise, the box contained 60 subunits, resulting in $\mbox{subunit concentration}=12 \mu$M. Assembly simulations were run at least 10 times for each parameter set, and each were concluded at $\tend=2\times10^8$ time steps. For each pseudoatom, $\gamma$ was assigned as its effective interaction diameter. VMD was used to visualize the model conformations \cite{Humphrey1996}. We calculated the hydrodynamic radius $\rh$ using HYDROPRO \cite{Ortega2011} as discussed in our previous work \cite{Perlmutter2014}.
\newpage

\subsection{Binding free energy estimates}
\label{sec:freeenergy}

We have previously calculated the free energy of subunit dimerization to be $\gss/\kt{=}5.0-1.5\ess$ \cite{Perlmutter2014,Perlmutter2013}. Briefly, simulations were set up with subunits which were limited to dimer formation, and the concentration of dimers was measured for varying $\ess$.  The free energy of binding along that interface is then $\gss/\kt =- \ln(c_\text{ss}/\kd)$ with standard state concentration $c_\text{ss}=1$ M and $\kd$ in molar units, and adjusted for the multiplicity of dimer conformations. The reduction in $\gss$ due to electrostatic repulsion between ARMs is $\sim0.5\kt$ at $\Csalt=100$mM.

We follow a similar strategy to calculate the binding free energy of the PS-PSR interaction. Here, we set up a simplified system containing a single trimeric PSR, composed of the PSR from three subunits, as well as the ARM and excluder pseudoatoms. The subunits (excepting the ARMs) are immobilized to prevent dissassembly. We then measured the relative concentration of PS bound and unbound states for a range of attraction strengths ($\eps$). The free energy of binding along that interface is then $\gps/\kt =- \ln(c_\text{ss}/\kd)$ with standard state concentration $c_\text{ss}=1$ M and $\kd$ in molar units. At $\Csalt=100$mM, the free energy is well fit by the linear expression $\gss/\kt = -1.3\ess - T\sbb$ where $T\sbb=2.3\kt$ (Fig.~\ref{psFreq}A). By this estimate our LA PS ($\eps=5\kt$) has $K_\text{D}\sim\mu$M. Note that this is the free energy of the PS binding to a complete, trimeric binding site; however, in many of our assembly simulations, the subunits do not form stable trimers without the PS. Therefore the PS-PSR interactions that occur during assembly often involve only one or two PSRs and non-optimal geometries.

Fig.~\ref{psFreq}B shows data from two related sets of simulations in which we measure the fraction of time a PS binds to PSRs in systems which contain a full length polyelectrolyte (575 segments at $\Csalt=100$mM) and 12 subunits.  In one case the 12 subunits are assembled into a complete capsid while in the other the subunits are adsorbed onto the polyelectrolyte but unassembled ($\ess$ was set to zero)., either as a completed capsid or unassembled, adsorbed subunits. Interestingly, while both curves are sigmoidal in shape, at binding strengths comparable to our LA PS ($\eps=5\kt$) PSs within complete capsids are nearly always bound to PSRs whereas they spend less than half their time bound to unassembled subunits. This difference arises because PSRs have ideal geometries within a capsid but are disordered in the unassembled subunits. Note that the binding probabilities measured in these simulations reflect a partitioning of PSs between specific binding to PSRs and nonspecific binding to subunit ARMs through electrostatics, whereas the free energies calculated in Fig.~\ref{psFreq}A reflect the partitioning between PSRs and solution.

\begin{figure}[h]
\centering{\includegraphics[width=0.75\columnwidth]{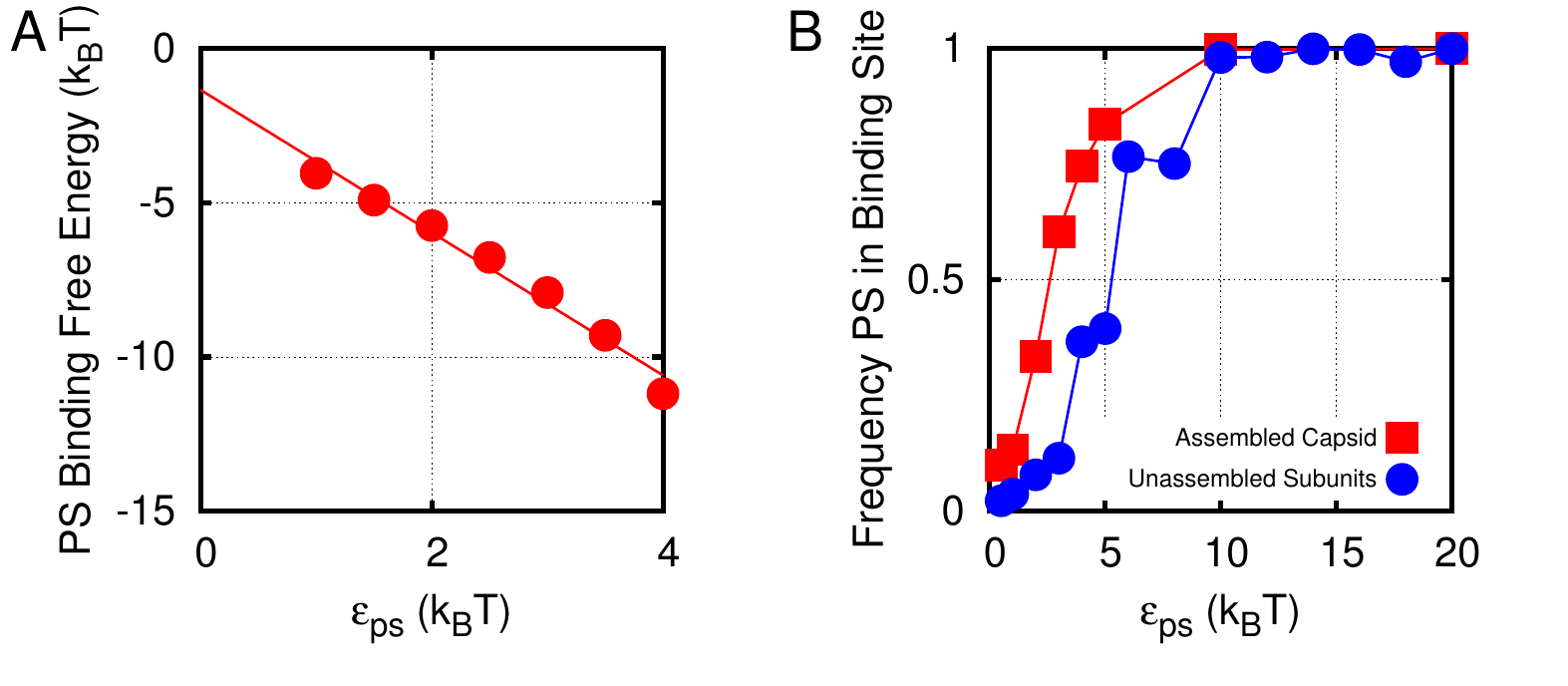}}
\caption{\textbf{(A)} Free energy of PS binding to a single complete, trimeric PSR binding site, as a function of PS-PRS interaction well depth $\eps$. The symbols indicate measured data points and the line shows a linear fit. \textbf{(B)} Fraction of time a PS is bound to PSRs as a function of $\eps$, measured within an assembled capsid (\textcolor{red}{$\blacksquare$} symbols) and in the presence of 12 adsorbed but unassembled subunits ({\large \textcolor{blue}{$\bullet$}} symbols). The polymer has 575 segments with one PS and $\Csalt=100$mM.}
\label{psFreq}
\end{figure}

\newpage
\section{Additional results and analysis}
\subsection{Effect of a single PS on assembly dynamics and specificity}
\label{sec:OnePS}
The fact that a single high affinity PS within a viral genome could promote specificity by functioning as a nucleation site has been considered for several decades (\eg \cite{Beckett1988})and the specificity conferred by a single PS was examined by in vitro experiments in which two species of heterologous RNA competed, one of which contained a single high affinity PS \cite{Beckett1988}.  These experiments identified modest selectivity ($\sim 2/3$) for the RNA with one PS.
To evaluate our results in the context of that experiment and to more broadly understand the limits of selectivity conferred by a single PS, we present simulations comparing non-cognate assembly with RNA containing a single HA PS located in the center of the polymer (as is the case for r17/MS2 \cite{Beckett1988}). We consider assembly at two salt concentrations ($\I=100,500mM$) in order to describe the effect of a PS under strong and weak electrostatic interactions respectively. For each salt concentration, we consider a range of subunit-subunit attraction strengths ($\ess/\kt \in [2,5]$ for $\Csalt=100$mM and ($\ess/\kt =6,7$ for $\Csalt=500$mM) over which assembly yields in the presence of the non-cognate RNA alone vary from zero to high (see Fig.~\ref{relativeYield}).  In the limit of low $\ess$ we observe disordered aggregates ($\Csalt=100mM, \ess=2\kt$) or failure to nucleate ($\Csalt=500mM, \ess=6\kt$) around the non-cognate RNA. The setup for these simulations is the same as for simulations presented in the main text.

In Fig.~\ref{growth1PS} the average size of the largest cluster of subunits is plotted as a function of time. For the parameters in which the non-cognate RNA triggers rapid assembly  ($\Csalt=100mM, \ess=4,5\kt$) incorporation of the PS does not substantially alter the time course of assembly. For cases in which assembly around the non-cognate RNA is slow ($\Csalt=100mM, \ess=3\kt$ and $\Csalt=500mM, \ess=7\kt$), the PS increases the assembly rate although long-time yields are similar. For cases in which the non-cognate RNA leads to no assembly on the investigated timescale ($\Csalt=100mM, \ess=2\kt$ and $\Csalt=500mM, \ess=6\kt$), the presence of a single PS increases the extent of assembly, but complete capsids do not form in the timescale considered here. As seen in Fig.~\ref{NumberPS}, additional PS are needed to induce efficient assembly at these parameters.

\begin{figure}[h]
\centering{\includegraphics[width=0.5\columnwidth]{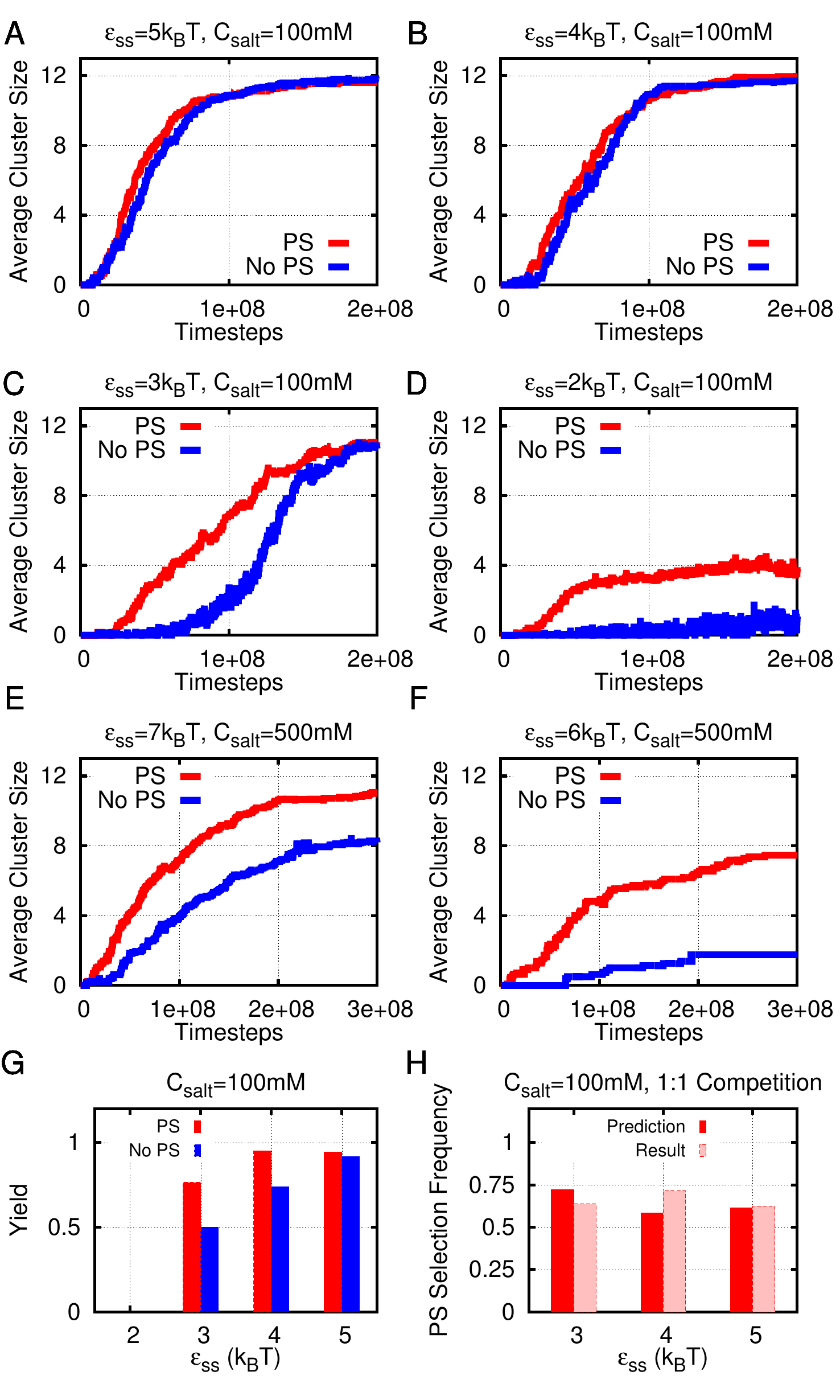}}
\caption{\textbf{(A-F)} Average cluster size as a function of simulation timestep with and without a single HA PS for varying salt concentration and subunit-subunit attraction. These results are the average of ten independent assembly simulations, each run using a single substrate and subunits at a concentration of 12 $\mu$M. \textbf{(G)} Yield of complete capsids from assembly simulations around a single non-cognate RNA or an RNA containing one HA PS, at several values of $\ess$ and $\Csalt=100mM$. \textbf{(H)} Comparison between specificity observed in explicit competition simulations and predicted by Eq.~{eq:P1}. The competition simulations each contained one non-cognate RNA, one cognate RNA containing one HA PS, and 60 subunits. The predictions use data from (A-G).}
\label{growth1PS}
\end{figure}

Figure~\ref{growth1PS}G shows the assembly yields around a single RNA with and without the PS for $\Csalt=100mM$. 20 simulations were run for each parameter value. As discussed above, for $\Csalt=100mM, \ess=5\kt$ assembly is robust without PS, and the effect of the PS on assembly is slight, while at $\ess=3,4\kt$ the increase in yield due to the PS is moderate. At $\ess=2\kt$ the presence of a single PS is not enough to allow for successful assembly and disordered aggregates are observed. In Figure~\ref{growth1PS}H we compare the selectivity predicted by Eq.~\ref{eq:P1} (using the data from Fig. ~\ref{growth1PS}A-G) against the results of explicit competition simulations. In these competition simulations, there are 60 subunits, one RNA containing a HA PS and one non-cognate RNA (thus $r_\text{ex}=1$). We find that the predicted selectivities match those from the explicit competition simulations quite closely, especially considering the limited number of independent simulations.  Moreover,  the finding that about $2/3$ of the assembled capsids contain the PS-RNA roughly agrees with the experimental observations of Beckett et al. \cite{Beckett1988}, in which incorporation of a single high affinity PS led to $\sim2/3$ selectivity under conditions where assembly around heterologous RNA was efficient.

\newpage

\subsection{Capsid and RNA subunit paths and conformational dynamics}
\label{sec:Stalling}

It has been proposed that PSs increase assembly rates by reducing the diversity of assembly pathways; essentially eliminating `dead end' pathways. To characterize the relationship between PSs and the ensemble of assembly pathways generated by our model, we evaluated the effects of PSs on the diversity of assembly intermediates, both from the perspective of arrangements of subunits in interacting clusters and RNA conformations.

\subsubsection{Partial capsid intermediate geometries}
We used two approaches to characterize the diversity of capsid intermediate geometries generated during assembly trajectories. In the first, we calculated the fraction of lowest-energy intermediates (a lowest energy intermediate contains the maximum possible number of subunit-subunit interactions for its size) for clusters with $3-11$ subunits. In the second, we categorized system configurations according to:  the number of clusters, the number of subunits in each cluster, and the number of bonds in each cluster. We then calculated the entropy $S$ of the distribution according to $S=-\sum_{\nu} \rho_\nu \log \rho_\nu$ with $\rho_\nu$ the relative probability of configuration $\nu$.

\begin{figure}[h]
\centering{\includegraphics[width=0.5\columnwidth]{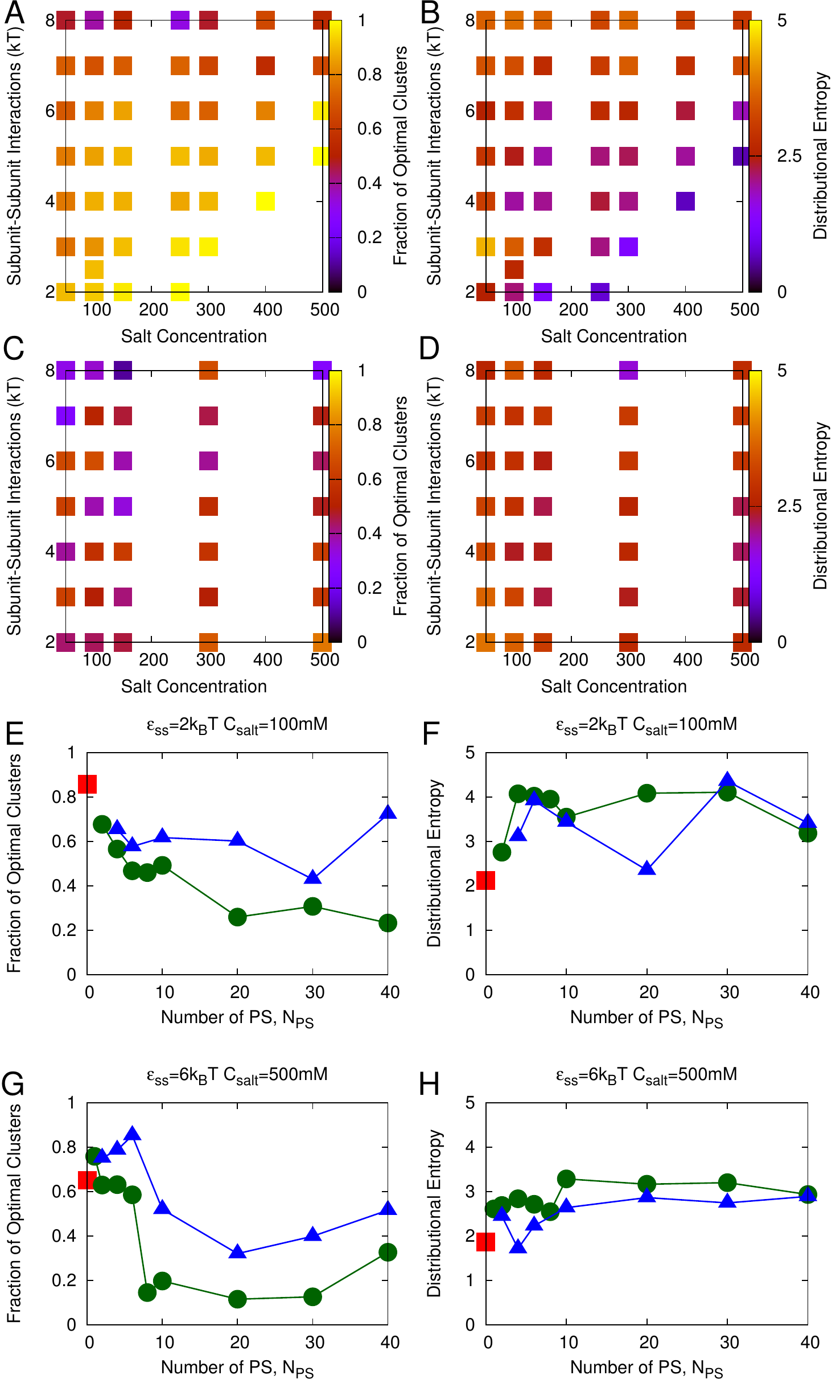}}
\caption{The fraction of assembly intermediates which contain the maximum number of subunit-subunit bonds (\textbf{(A,C,E,G)} ) and the entropy of the distribution of intermediate configurations (\textbf{(B,D,F,H)} ).  Results are shown as a function of $\ess$ and $\Csalt$ for a non-cognate RNA \textbf{(A,B)}  and the Combo PS sequence \textbf{(C,D)} .  Results are shown as a function of the number of PSs $\nps$, with $\ess=2\kt$ and $\Csalt=100$mM \textbf{(E,F)}  and $\ess=6\kt$ and $\Csalt=500$mM \textbf{(G,H)} . In \textbf{(E-H)} the  RNA contained either $\nps$ HA PSs ({\large\textcolor{OliveGreen}{$\bullet$}} symbols) or 1 HA PS and $\nps$ LA PSs (\textcolor{blue}{$\blacktriangle$} symbols). Results for the non-cognate RNA (no PSs) are indicated by \textcolor{red}{$\blacksquare$} symbols.
}
\label{subPath}
\end{figure}

Fig.~\ref{subPath}A,B shows the results for non-cognate RNA assembly, as a function of salt concentration and subunit-subunit interaction. We see that weak subunit-subunit interactions and high salt concentrations lead to a high fraction of optimal intermediate geometries and consequently a low entropy of the distribution of intermediate configurations. As $\ess$ increases or $\Csalt$ decreases the distribution of intermediate configurations widens, indicating more diverse assembly pathways.
The results are consistent with a previous study of uniform polyelectrolytes \cite{Perlmutter2014}, in which we found that high salt and moderate subunit-subunit interactions lead to an `ordered' nucleation-and-growth assembly which proceeds through a series of well-formed intermediates.  Stronger interactions (either electrostatic or subunit-subunit) tend to promote the formation of multiple clusters and stabilize non-optimal configurations.

The results with the Combo PS sequence (Fig.~\ref{subPath}C,D) show that PSs tend to increase the diversity of intermediate configurations and prevalence of non-optimal cluster geometries. The effect is particularly noticeable under conditions where the underlying interactions are weak (high salt, low $\ess$).  Although the results as a function of varying $\nps$ (Fig.~\ref{subPath}E-H) are noisy, we generally observe an increase in the width of the distribution and more non-optimal clusters as more PSs are added, and the HA PSs have a stronger effect than the LA PSs.

In contrast to the general observation that adding PSs leads to a wider diversity of intermediate geometries, we find that PSs narrow the distribution of intermediate geometries under parameters where electrostatics and subunit-subunit interactions are both strong (\eg $\Csalt=50mM, \ess=7\kt$). This effect is significant; as discussed in the main text (section~\ref{sec:results}), PSs increase assembly yields in this regime even though the nonspecific interactions are sufficiently strong to promote assembly (Fig.~\ref{relativeYield}).  To clarify the mechanism by which PSs influence pathways in this regime, we separately characterized their effect on intermediate geometries and distribution of intermediates.  In particular, we calculate the average deviation from the ground state for clusters with $n$ subunits, given by \cite{Hagan2011}
\begin{equation}
\bad(n)  = \langle \delta_{n(c),n} \left[B^\mathrm{gs}(n)  - B(c)\right] \rangle_c.
\label{eq:badness}
\end{equation}
where $B(c)$ is the number of subunit-subunit interactions for configuration $c$, $n(c)$ is the number of subunits in configuration $c$, $B^\mathrm{gs}(n)$ is the number of bonds in the minimum-energy configuration of $n$ subunits, and the average is taken over all configurations weighted by their frequency of appearance in assembly trajectories.  The quantity $\bad$ is shown as a function of intermediate size in Fig.~\ref{fig:highHigh}A for a non-cognate RNA and the Combo PS sequence.  We see that the PSs {\bf increase} the frequency of deviations from the ground state cluster, consistent with other regions of parameter space.  However, the average number of clusters  (Fig.~\ref{fig:highHigh}B) significantly decreases in the presence of PSs.  We find that the HA PS in the Combo sequence promotes rapid nucleation of a partial capsid, which tends to complete assembly before other nucleation events occur. In contrast, multiple nucleation events are common on the uniform polyelectrolytes. As noted in the main text, avoiding multiple nucleation events appears to be the mechanism by which PSs increase yields and confer specificity in this regime of relatively strong nonspecific interactions.

\begin{figure}[h]
\centering{\includegraphics[width=0.4\columnwidth]{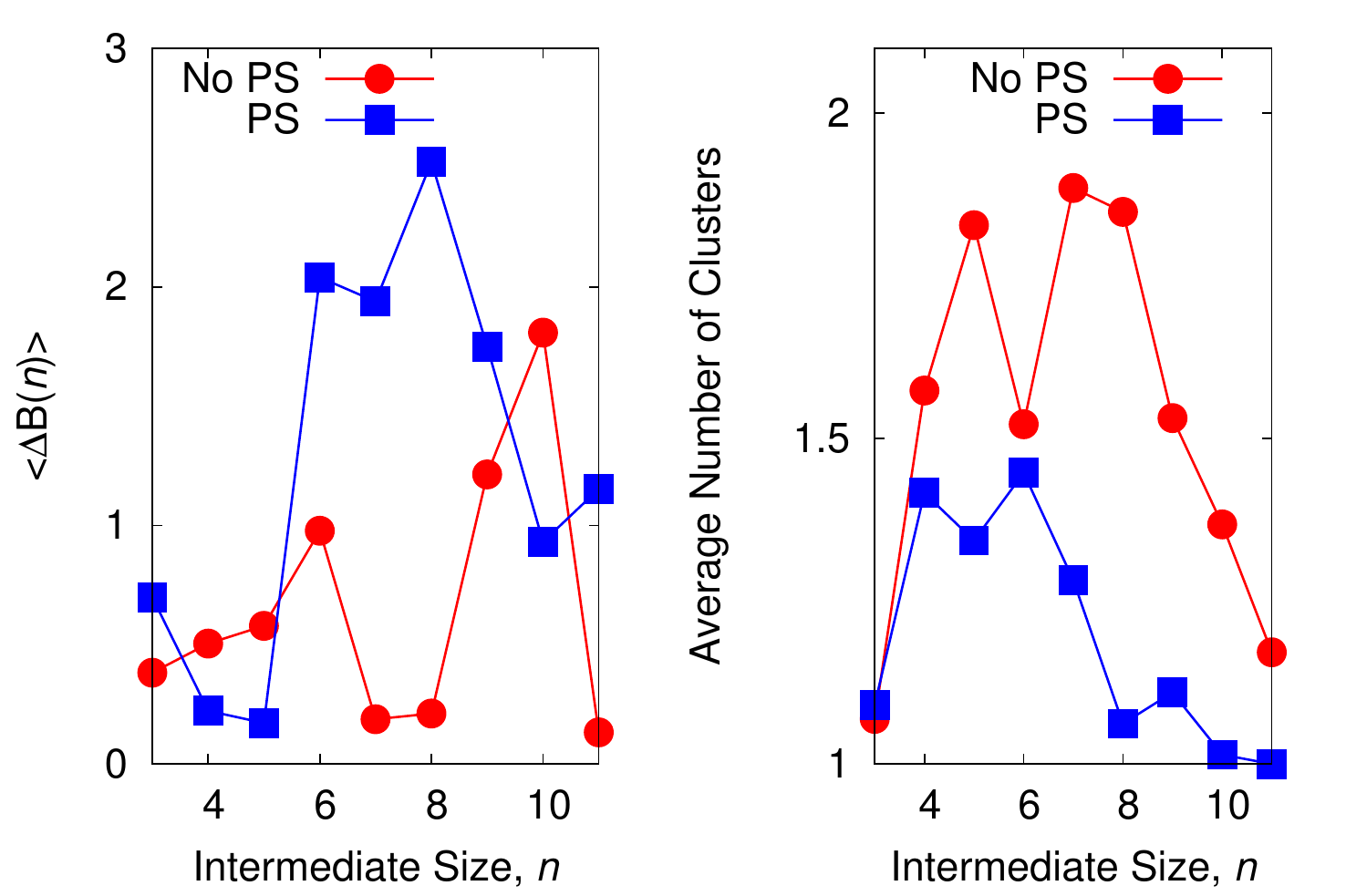}}
\caption{{bf (A)} Average deviation in number of interactions $\bad$ from the ground state configuration as a function of  intermediate size  $n$, averaged over assembly trajectories with a uniform polyelectrolyte (\textcolor{red}{$\bullet$} symbols) or the Combo PS sequence (\textcolor{blue}{$\blacksquare$} symbols). {\bf (B)} Number of clusters as a function of intermediate size $n$ averaged over assembly trajectories. Simulation parameters were $\Csalt=50mM, \ess=7\kt$, with the cognate RNA containing the Combo PS sequence 1 HA PS + 25 LA PS.}
\label{fig:highHigh}
\end{figure}

\subsubsection{RNA conformational statistics and dynamics}
To simplify characterization of the ensemble of RNA conformations and its dynamics, we categorize RNA conformations according to the sequence of capsid vertices with which the substrate interacts.  This definition neglects fluctuations of RNA segments between vertices, and thus most closely reflects RNA conformations in simulations with high salt, large $\eps$, and $\nps\approx20$, for which the PS-PSR binding is the dominant interaction. Under parameters with fewer PSs and stronger electrostatics, although we observe enhanced polyelectrolyte density in the vicinity of capsid vertices \cite{Perlmutter2013}, this definition does not completely reflect the diversity of RNA configurations.

\begin{figure}[h]
\centering{\includegraphics[width=0.35\columnwidth]{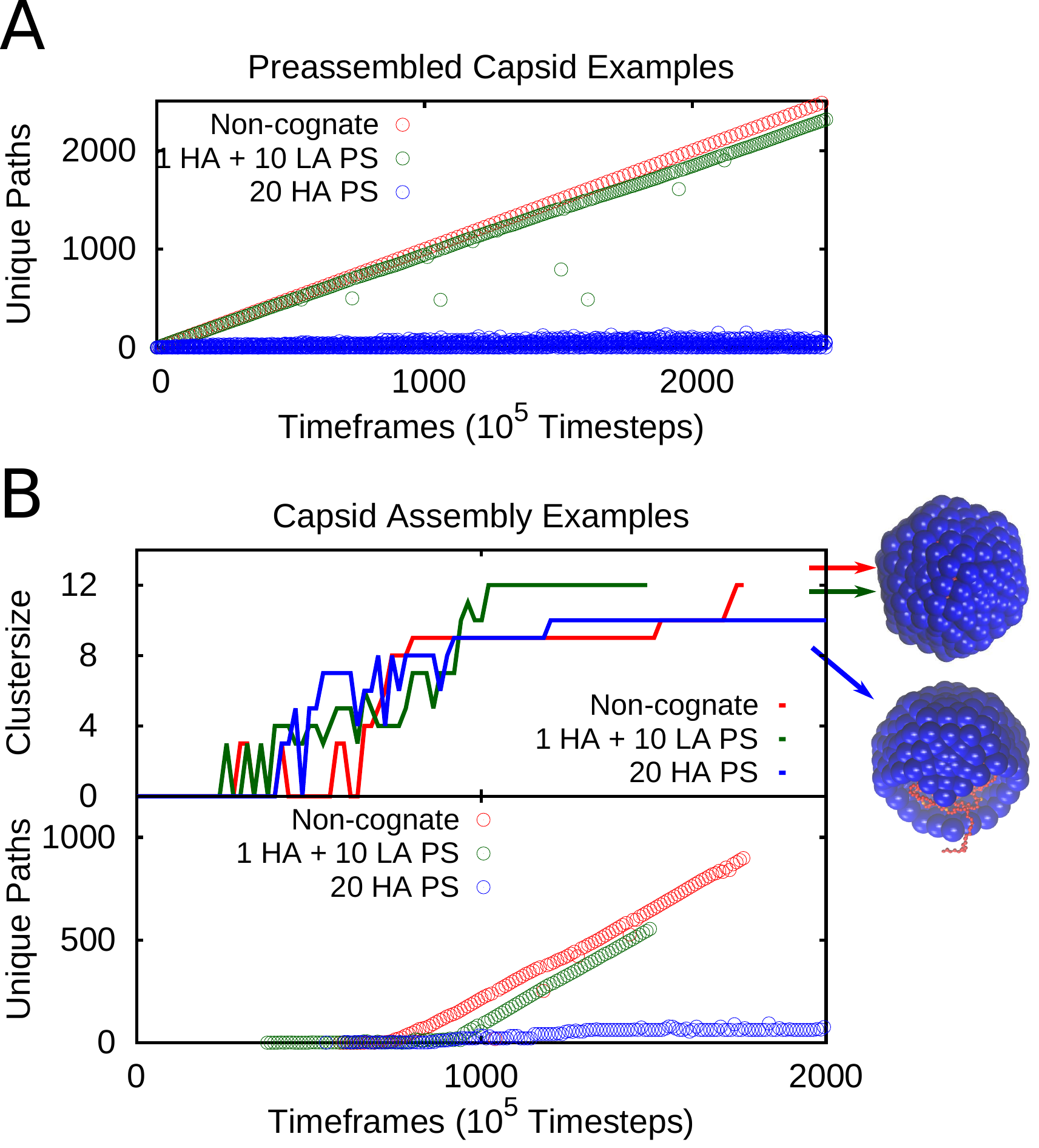}}
\caption{Examples showing RNA path exploration within preassembled capsids (A) and for assembly simulations (B). At each step we determine the polymer path, and if unique assign it a new index. For the Non-cognate and `1 HA + 10 LA PS' cases, a unique path occurs at nearly every frame, whereas the `20 HA PS' case is limited to a very small set of paths. We observe that this restriction in RNA dynamics corresponds with stalled assembly. These simulations are run at $\Csalt=100mM$, and $\ess=5\kt$ for the non-cognate or $\ess=2\kt$ for the PS containing polymers.}
\label{pathExample}
\end{figure}

Fig.~\ref{pathExample} shows the frequency with which new RNA conformations (according to the above definition) are generated during simulations of RNA within a completely assembled capsid (Fig.~\ref{pathExample}A) and within an assembling capsid (Fig.~\ref{pathExample}B). For the non-cognate and LA PS cognate sequences, the RNA adopts a new path at almost every window, while the RNA with 20 HA PSs undergoes restricted dynamics, rarely transitioning to a new conformation. Comparison of Fig.~\ref{pathExample} and Fig.~\ref{NumberPS} suggests that frozen RNA dynamics tend to give rise to stalled assembly trajectories; indeed, assembly stalls at 10 subunits for the 20 HA PS sequence (Fig.~\ref{pathExample}B upper frame). These simulations are run at $\Csalt=100$mM.

\begin{figure}[h]
\centering{\includegraphics[width=0.6\columnwidth]{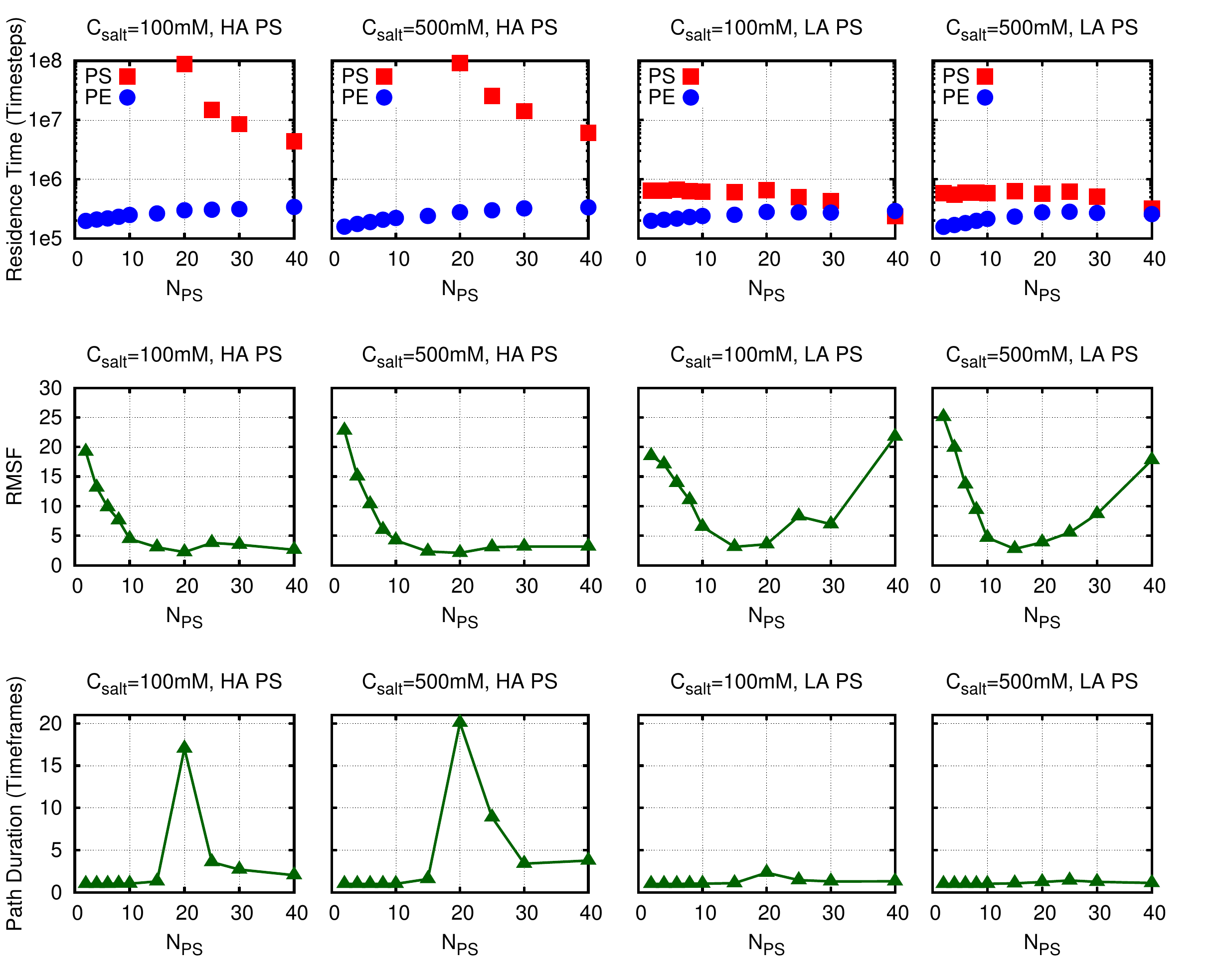}}
\caption{Different measures of RNA dynamics within an assembled capsid. The first row presents the average residence time for PS-binding site interactions and Polyelectrolyte-ARM interactions. The second row presents the average root mean squared fluctuations (averaged over time and RNA segments). The third row presents the average duration of RNA paths. These simulations are performed for a preassembled capsid containing a single RNA of optimal length with the specified $\nps$, with $\eps=20\kt$ for HA PS and $\eps=50\kt$ for LA PS.}
\label{polDynamics}
\end{figure}

Previous simulations suggested that polymer-mediated capsid assembly is enhanced by cooperative polymer-subunit motions and by `sliding' of adsorbed subunits along the polymer \cite{Kivenson2010,Hu2007,Elrad2010}. To evaluate the extent to which adsorbed subunits can rearrange, in Figure~\ref{polDynamics} we quantify  the average residence time for the PS-PSR interaction (PS) and the Polyelectrolyte-ARM interaction (PE) within a completed capsid for a polymer with a varying number of HA or LA PS ($\eps=20, 5\kt$). As expected, PS exchange is slow in comparison to exchange of PE interactions, with exchange of HA PSs significantly slower than exchange of LA PSs.  (For HA PSs, exchange often occurs on longer timescales than our simulation times.) However, excess PSs ($\nps>20$) appear to facilitate exchange.

In the second row of Figure~\ref{polDynamics} we quantify the root mean squared fluctuations (RMSF) of the RNA within the capsid. We observe that increasing the number of PSs reduces the overall RNA dynamics. However, introduction of excess LA PSs leads to a rebound in the dynamics, further emphasizing the relationship between excess PSs and RNA rearrangements. In the third row, we quantify the average duration of paths within the capsid, which emphasizes the restriction of dynamics at $\nps=20$ HA PS.

It is interesting to compare these different measures of dynamics in Figure~\ref{polDynamics}. For example, the first and third rows indicate an increase in dynamics for $\nps>20$, however this is not captured by the RMSF. We infer that the exchange of HA PS, even at excess $\nps$, is a slower process, and thus their effect is obscured by other measures of RNA dynamics. However, this slow motion is still functionally relevant, as observed in the increase in yield with increasing HA $\nps$.

\newpage

\subsection{Effect of PS on equilibrium encapsidation and substrate length}
\label{sec:eq}
In this section we describe the effect of PSs on the thermodynamically optimal RNA length for encapsidation $\Lopt$, defined as the length which minimizes the free energy of the RNA-capsid complex \cite{Perlmutter2013}. We found previously that $\Lopt$ corresponds to the length which optimizes finite-time assembly yields, at least within measured parameter ranges \cite{Perlmutter2013}.  We also found that, when the model was adapted to match features of specific viruses, namely the interior volume of the capsid and the length and charge of the ARM amino acids, the calculated values of $\Lopt$ closely agree with the genome length for those viruses when effects of base pairing were included in the model \cite{Perlmutter2013}.

Here, we present the effect of PSs on $\Lopt$.   We calculate $\Lopt$ using the same protocol as in \cite{Perlmutter2013, Perlmutter2014}; a very long RNA strand is placed in a preassembled capsid, with a small section of the capsid rendered permeable to the RNA.  We then perform dynamics during which the RNA rearranges within the capsid and partially extrudes from the permeable section of the capsid. Once the system has equilibrated, $\Lopt$ is calculated as the average number of RNA segments remaining within the capsid, measured over a period of ($5*10^{7}$ timesteps).  Values of  $\Lopt$ measured by this protocol were found to closely agree with the values which minimized the free energy calculated using the Widom test-particle method \cite{Widom1963} as extended to calculate polymer residual chemical potentials \cite{Kumar1991, Elrad2010,Perlmutter2013}.

The dependence of $\Lopt$ on the PS-PSR binding strength $\eps$  is shown in Fig.~\ref{equilibrium} for RNA with varying numbers of PSs.  We report the frequency of PSs, defined as the number of polyelectrolyte segments between each PS (which are uniformly spaced), since the number of encapsidated PSs can depend on $\Lopt$.  We see that, while a single PS has essentially no effect on $\Lopt$, increasing the PS strength and frequency can significantly increase $\Lopt$.  Note that since the single PS has essentially no effect on $\Lopt$, the optimal length for the Combo sequence with 1 HA PS and 25 LA PSs is approximately 628 (relative to 575 for the noncognate).

\begin{figure}[h]
\includegraphics[width=0.75\columnwidth]{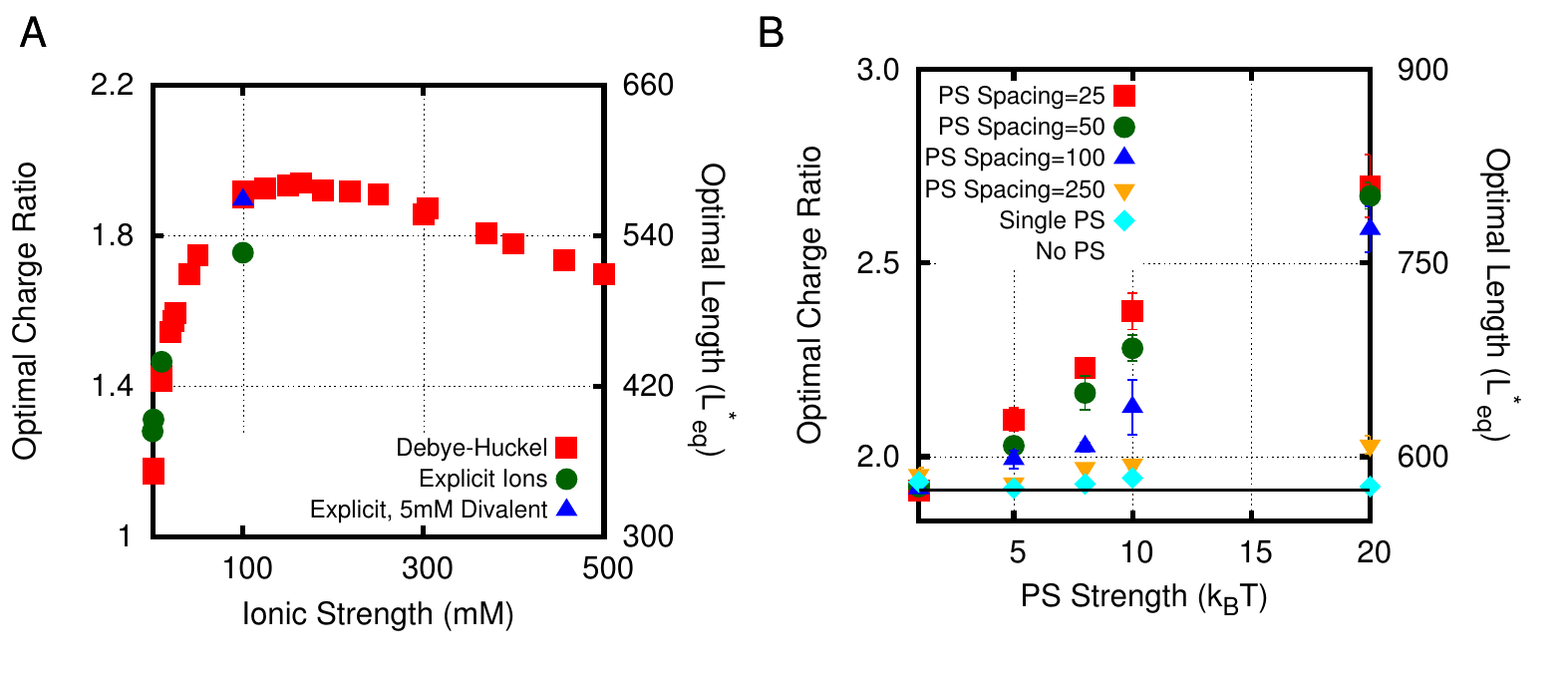}
\caption{\textbf{(A)} The thermodynamic optimal RNA length $\Lopt$ is shown as a function of $\Csalt$ with monovalent salt for a uniform polyelectrolyte (the non-cognate RNA). \textbf{(B)} Dependence of $\Lopt$ on PS-PSR binding strength  $\eps$, for RNAs with uniformly spaced PSs, with indicated numbers of RNA segments between each PS.  The salt concentration is $\Csalt=100$mM.}
\label{equilibrium}
\end{figure}

We previously found that dynamical assembly of complete capsids around uniform polyelectrolytes occurs  only for polyelectrolyte lengths within about 10\% of $\Lopt$. In Fig.~\ref{length} we show the yield of well-formed capsids at the end of long but finite-time dynamical simulations for uniform polyelectrolyte and the Combo PS sequence. We see that the PSs slightly increase the optimal length, to approximately the same value as the thermodynamic $\Lopt$.  While the distribution of yields as a function of polyelectrolyte lengths is slightly broader for the PS sequence than the uniform polyelectrolyte, variations are still limited to $\sim 10\%$.  This result corroborates the observation of Fig.~\ref{equilibrium}, that the increase in $\Lopt$ for the LA PSs is relatively small, suggesting that the optimal RNA length does not differ significantly between non-cognate and cognate RNAs. This result is consistent with the observation values of $\Lopt$ from a model which did not account for PSs agreed with actual genome lengths.

\begin{figure}[h]
\centering{\includegraphics[width=0.35\columnwidth]{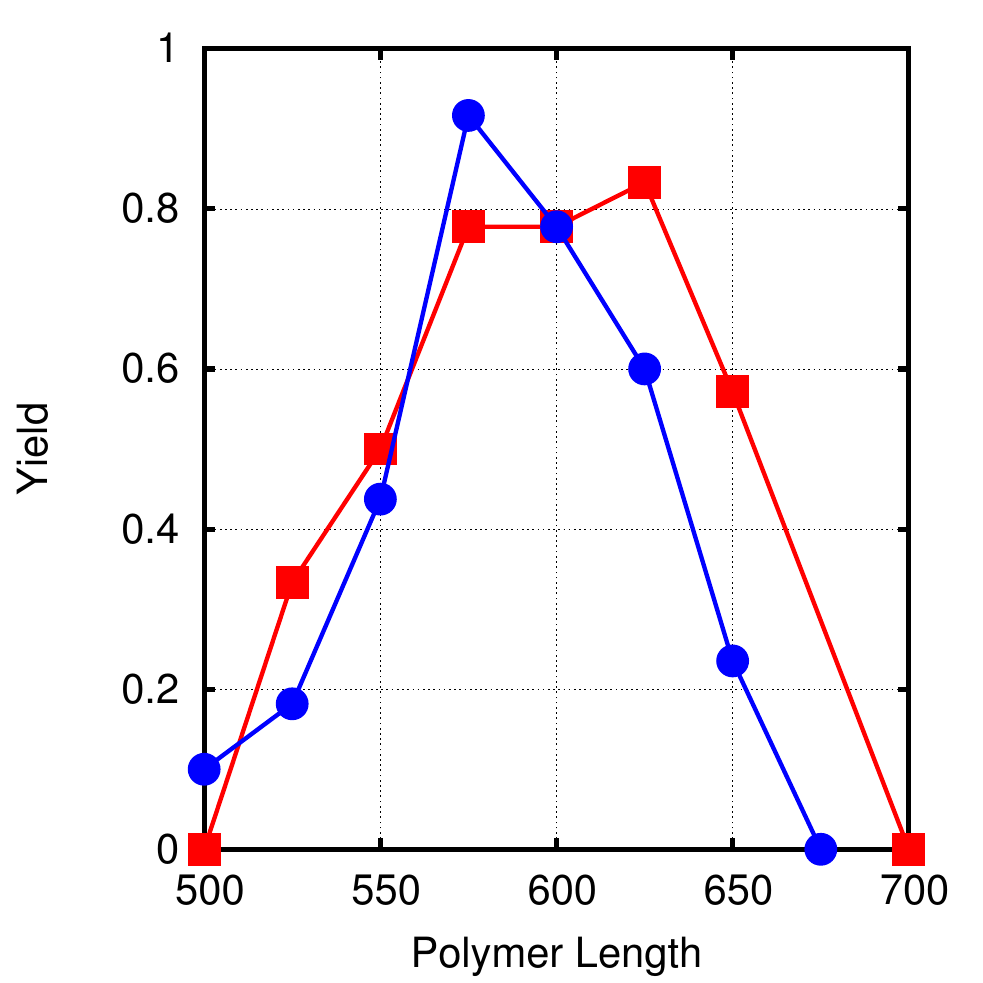}}
\caption{Yield of well-formed capsids in dynamical assembly simulations as a function of length of a polyelectrolyte with no PSs ({\large \textcolor{blue}{$\bullet$}} symbols) or 1 HA + 20 LA PSs (\textcolor{red}{$\blacksquare$} symbols).  Parameters are $\ess=5\kt$ for the uniform polyelectrolyte, $\ess=2\kt$ for the PS sequence, and $\Csalt=100$mM in both cases.}
\label{length}
\end{figure}

\subsection{Dependence of RNA path properties on ARM length}
Previous simulations showed that $\Lopt$ depends on charge and excluded volume of the capsid, as determined by the ARM charge, ARM length and capsid interior volume \cite{Perlmutter2013}. Here we present results from simulations of complete capsids with varying ARM charge, encapsidating RNA with $\nps=20$ LA PS. The ARM charge determines the optimal length of the substrate \cite{Perlmutter2013}, which determines the spacing between the PS and the geometry of the path within the capsid. Figure~\ref{singleness} is a histogram showing the distribution of step lengths for the polymer path (sec~\ref{sec:Stalling}); i.e. a step length of $1$ corresponds to stepping to a nearest neighbor vertex, while a step of length $2$ corresponds to skipping one vertex. Interestingly, this distribution behaves non-monotonically; for short ARMs ($2,3$) stepping to the nearest neighbor is strongly favored, for $ARM=4$ there is a clear shift towards a separation of $2$, which then is reversed for $ARM=5$. Our analysis here suggests that the structure of the capsid protein-substrate complex and the ability of the substrate to promote assembly is determined by many factors, including the capsid geometry and charge, substrate length and charge, and PS number, strength, and spacing.

\begin{figure}[h]
\centering{\includegraphics[width=0.45\columnwidth]{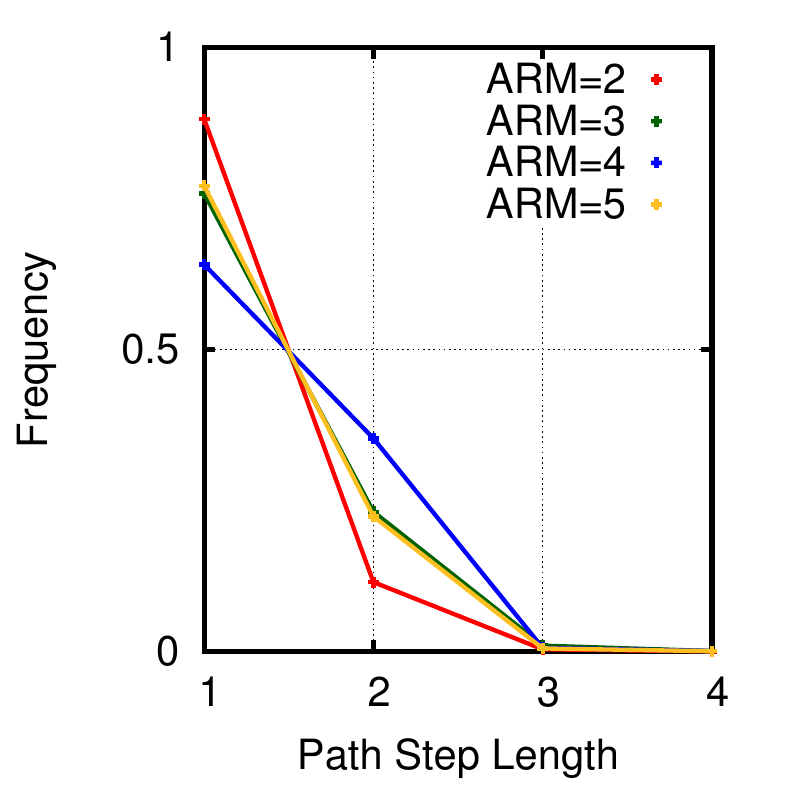}}
\caption{Histogram describing the conformation of encapsidated RNA with $\nps=20$ PSs for varying ARM lengths. Here the X-axis is distance between vertices traced in the polymer path; a step length of $1$ indicates a step to a nearest-neighbor vertex, while a step length of $2$ indicates a step which skips a vertex. For each ARM length, the optimal RNA length was used:  ARM=2: 350 segments, ARM=3: 428 segments, ARM=4: 504 segments, ARM=5: 575 segments. This changes the PS spacing, and the frequency of step length.}
\label{singleness}
\end{figure}

\subsection{Simulation Movies}

Movie 1 - Direct competition simulation between a noncognate (red) and cognate (magenta) with one HA PS. Simulation conditions: $\Csalt=100mM, \ess=5\kt, r_\text{ex}=1$, excess subunits.
\bigskip

Movie 2 - Direct competition simulation between a noncognate (red) and cognate (magenta) using the `Combo' PS sequence $\nps=20$. Simulation conditions: $\Csalt=100mM, \ess=2\kt, r_\text{ex}=1$, excess subunits.
\bigskip

Movie 3 - Direct competition simulation between a noncognate (red) and cognate (magenta) using the `Combo' PS sequence $\nps=30$. Simulation conditions: $\Csalt=500mM, \ess=6\kt, r_\text{ex}=1$, limiting subunits.

\end{document}